\documentclass[preprint,journal]{vgtc}            

\onlineid{0}

\preprinttext{To appear in IEEE Transactions on Visualization and Computer Graphics. doi: 10.1109/TVCG.2024.3456319}

\ieeedoi{10.1109/TVCG.2024.3456319}

\vgtccategory{Research}

\vgtcpapertype{application/design study}

\title{User Experience of Visualizations in Motion:\\ A Case Study and Design Considerations}

\author{%
  \authororcid{Lijie Yao}{0000-0002-4208-5140}, 
  \authororcid{Federica Bucchieri}{0009-0009-6398-0660},
  \authororcid{Victoria McArthur}{0000-0002-6650-6886},
  \authororcid{Anastasia Bezerianos}{0000-0002-7142-2548}, and
  \authororcid{Petra Isenberg}{0000-0002-2948-6417} 
}

\authorfooter{
  \item
  	L. Yao is with Xi'an Jiaotong-Liverpool University, China, and Universit{\'e} Paris-Saclay, CNRS, Inria, France. E-mail: yaolijie0219@gmail.com.
   \item
   A. Bezerianos, and P. Isenberg are with Universit{\'e} Paris-Saclay, CNRS, Inria, France.
  	E-mail: anastasia.bezerianos@universite-paris-saclay.fr, petra.isenberg@inria.fr.
  \item
  	F. Bucchieri was with Universit{\'e} Paris-Saclay, CNRS, Inria, France.
  	E-mail: federicabucchieri@gmail.com.

  \item V. McArthur is with Carleton University.
  	E-mail: victoria.mcarthur@carleton.ca.
}

\abstract{
We present a systematic review, an empirical study, and a first set of considerations for designing visualizations in motion, derived from a concrete scenario in which these visualizations were used to support a primary task.
In practice, when viewers are confronted with embedded visualizations, they often have to focus on a primary task and can only quickly glance at a visualization showing rich, often dynamically updated, information. As such, the visualizations must be designed so as not to distract from the primary task, while at the same time being readable and useful for aiding the primary task. For example, in games, players who are engaged in a battle have to look at their enemies but also read the remaining health of their own game character from the health bar over their character's head. Many trade-offs are possible in the design of embedded visualizations in such dynamic scenarios, which we explore in-depth in this paper with a focus on user experience. We use video games as an example of an application context with a rich existing set of visualizations in motion. We begin our work with a systematic review of in-game visualizations in motion. Next, we conduct an empirical user study to investigate how different embedded visualizations in motion designs impact user experience. We conclude with a set of considerations and trade-offs for designing visualizations in motion more broadly as derived from what we learned about video games. All supplemental materials of this paper are available at \href{https://osf.io/3v8wm/}{osf.io/3v8wm/}.
}


\graphicspath{{figs/}{figures/}{pictures/}{images/}{./}} 
\usepackage{tabu}                      
\usepackage{booktabs}                  
\usepackage{lipsum}                    
\usepackage{mwe}                       
\usepackage[export]{adjustbox}

\usepackage{mathptmx}                  
\usepackage[dvipsnames,svgnames,xcdraw, table]{xcolor}
\usepackage{cite}

\usepackage{ccicons}
\usepackage{xspace}
\usepackage{amsmath}
\usepackage{colortbl}
\usepackage{microtype}
\usepackage{comment}
\usepackage{expdlist}
\usepackage{flushend}
\usepackage[super]{nth}
\usepackage{subfig}
\usepackage{graphicx}
\usepackage{array}
\usepackage{amsmath}
\usepackage{float}
\usepackage{subcaption}
\usepackage{enumitem}
\usepackage{tabu}
\usepackage[compact]{titlesec}

\titlespacing*{\section}
{0pt}{1.5ex plus 1.3ex minus .2ex}{0.5ex plus .2ex}
\titlespacing*{\subsection}
{0pt}{1ex plus 1.3ex minus .2ex}{0.5ex plus .2ex}

\usepackage[skip=0pt]{caption}

\newcommand{\ts}{\textsuperscript}

\usepackage{siunitx}

\usepackage{xspace}
\newcommand{\ie}{i.\,e.\@\xspace}
\newcommand{\eg}{e.\,g.\@\xspace}
\newcommand{\etal}{et al.\@\xspace}

\definecolor{upsaclay}{rgb}{0.39,0.0,0.23}
\definecolor{bluePoli}{rgb}{0.39,0.0,0.23}
\definecolor{nicered}{rgb}{0.39,0.0,0.23}

\makeatletter
\DeclareRobustCommand\onedot{\futurelet\@let@token\@onedot}
\def\@onedot{\ifx\@let@token.\else.\null\fi\xspace}
\def\eg{\emph{e.g}\onedot} 
\def\ie{\emph{i.e}\onedot}

\def\etal{\emph{et al}\onedot}
\makeatother

\newlength{\myLength}	
\newlength{\mytextsize}

\usepackage{tikz}
\newcommand{\percentagebar}[1]{\begin{tikzpicture}[baseline = .15\mytextsize]%
\draw[fill=white, thin,opacity=1] [yshift=1pt] (0,0) rectangle (\myLength,.6\mytextsize);
\draw[fill=black, thin,opacity=0.7] [yshift=1pt] (0,0) rectangle (#1 \myLength,.6\mytextsize);
\end{tikzpicture}
}

\usepackage{wasysym}
\newcommand{\ageRange}{\includegraphics[height = 3ex, valign=c]{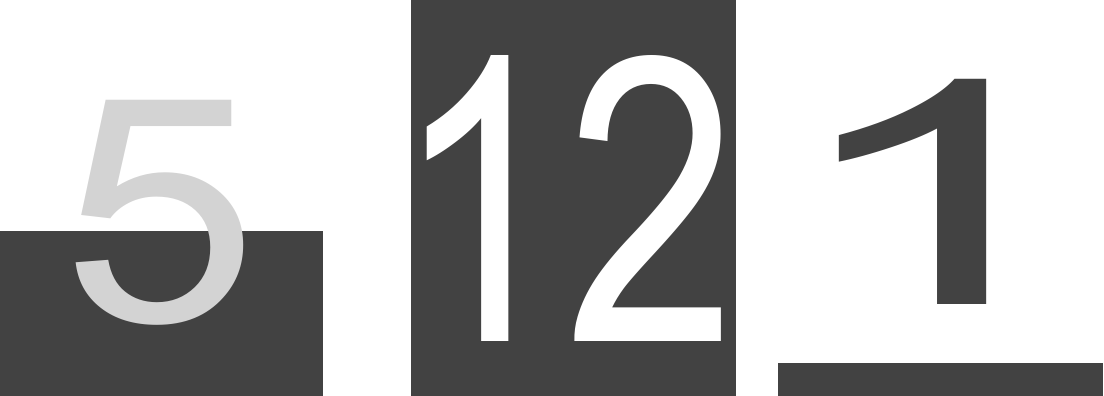}}

\newcommand{\barS}{the \texttt{non-integrated design}}
\newcommand{\textureS}{the \texttt{fully-integrated design}}
\newcommand{\donutS}{the \texttt{partial-match design}}

\newcommand{\BarS}{The \texttt{non-integrated design}}
\newcommand{\TextureS}{The \texttt{fully-integrated design}}
\newcommand{\DonutS}{The \texttt{partial-match design}}

\newcommand{\barIcon}{\includegraphics[height = 3ex, valign=c]{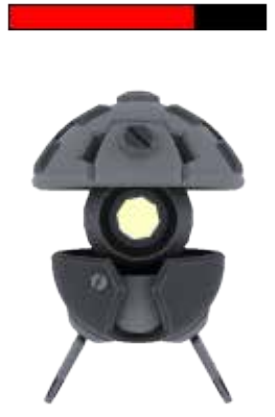}}
\newcommand{\textureIcon}{\includegraphics[height = 3ex, valign=c]{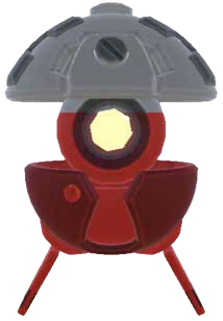}}
\newcommand{\donutIcon}{\includegraphics[height = 3ex, valign=c]{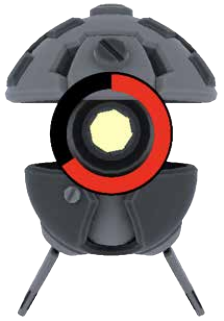}}

\usepackage{tabularx}
\newcolumntype{L}[1]{>{\raggedright\arraybackslash}p{#1}}
\newcolumntype{C}[1]{>{\centering\arraybackslash}p{#1}}
\newcolumntype{R}[1]{>{\raggedleft\arraybackslash}p{#1}}

\usepackage{multirow}
\newlength{\picturewidth}
\newlength{\pictureheight}

\usepackage{calc}
\usepackage{fp}
\newcommand{\maxnum}{18}
\newlength\maxlen
\newcommand{\databar}[2][upsaclay]{%
  \settowidth{\maxlen}{\maxnum}%
  \addtolength{\maxlen}{\tabcolsep}%
  \FPeval\result{round(#2/\maxnum:3)}%
  \raisebox{0pt}[0pt][0pt]{\rlap{\color{#1}\hspace*{-.7\tabcolsep}\rule[-.20\ht\strutbox]{\result\maxlen}{1.20\ht\strutbox}}}%
  \makebox[\dimexpr\maxlen-\tabcolsep][r]{#2}%
}

\definecolor{saclayPurple}{RGB}{90,6,48}
\definecolor{saclayPurpleLight}{RGB}{212,159,185}
\definecolor{isoorange}{RGB}{252,182,122}
\definecolor{isogreen}{HTML}{a8ddb5}
\definecolor{isoblue}{RGB}{110,208,255}

\makeatletter
\DeclareRobustCommand\onedot{\futurelet\@let@token\@onedot}
\def\@onedot{\ifx\@let@token.\else.\null\fi\xspace}
\def\eg{\emph{e.g}\onedot} 
\def\ie{\emph{i.e}\onedot}

\def\etal{\emph{et al}\onedot}
\makeatother

\newcommand{\revisionVIS}[1]{\textcolor{BrickRed}{#1}}

\begin{document}



\firstsection{Introduction}
\maketitle
On interactive media, visualizations can be embedded and move with their data referents, such as people, objects, or places the represented data refers to.
Examples include visual representations of player performance attached to and moving with athletes in sports videos \cite{Chen:2023:iBall, Yao:2022:VisForSwimming, Chen:2021:VisCommentator}, visualizations attached to moving entities in augmented reality \cite{Lin:2022:Omnioculars, Lin:2020:SportsXR, Reipschlager:2021:ARinteraction, Lee:2022:DesignSpace, Lee:2024:DesignSpace}, simple graphics displayed on fitness trackers to show real-time exercising data \cite{Blascheck:2023:ParttoWhole, Islam:2022:SleepData, Yao:2022:FitnessTrackers, Blascheck:2019:Glanceable}, or the in-game visualizations that move with and are embedded very close to game characters \cite{VisInGame:2008, TowardVisualizationforGames, Martinez:2012:StudyOfColorInGameEngagement, Milam:2011:cameramotiongame}. Yao \etal\ call visualizations that have a relative movement relationship to their viewers \emph{``visualizations in motion''} \cite{Yao:2022:VisinMotion}. 

In the above scenarios, viewers generally have a primary task, such as watching a race, tracking an object, playing a game, or exercising.
In some of them, visualizations in motion are used as an additional information provider but may not be required for viewers' primary tasks. For example, the absence of visualizations in a swimming broadcast will not change the race result.  However, in some cases, visualizations in motion may serve more than auxiliary functions and can participate in the viewers' primary task.
For example, when running with fitness trackers,  seeing real-time information about speed from their smartwatches may encourage runners to run faster. Similar cases exist in video games. The in-game visualizations that are embedded close to and move with their game characters, like health bars \cite{HealthBarWiki}, inform players about the characters' real-time health status. In such cases, visualizations in motion may influence actions people might take, like deciding which is a weaker enemy to attack.

The creation of embedded visualization in motion designs comes with various trade-offs to consider. These trade-offs include visualization design parameters (\eg, representation, size, color, transparency), where to embed the visualization (\eg, around or  into its data referent), or what data to display (\eg, how many data points, dimension, with dynamic updates or not).  
These design choices may affect viewers' primary tasks and their user experience, especially in contexts where viewers can only intermittently check the embedded visualizations. 

To understand how design differences can affect the user experience of visualizations in motion, we selected video games as an example application context for our research for the following reasons: (a) \emph{Visualization prevalence:} Video games produce rich, dynamic datasets during game-play that are often visualized to enhance the player experience. Examples include health bars, navigation aids, ammunition count, or team affiliation. (b) \emph{Primary tasks}: In-game visualizations often need to be read at a glance while the player is focused on a primary task, such as fulfilling a game mission. (c) \emph{Motion involvement and rich context:} Frequently, in-game visualizations are in motion on the screen due to camera changes or because the visualizations are attached to moving game elements. In-game visualizations are also often restricted in size. Their aesthetics need to fit a game world, and their colors need to be readable across a variety of background textures.  
We first conduct a systematic review of in-game visualizations in motion to investigate what designs are used in practice and how design challenges have been addressed in the past. Next, we run an empirical study with gaming enthusiasts on visualizations in motion embedded in a video game --- \emph{RobotLife}. After analyzing the gaming performance and the design preference, we propose a first set of design considerations.

Note that our goal is not to provide recommendations about how to design visualizations for video games, but rather to treat video games as a test context to explore how to design visualizations in motion. 
Our work contributes to the design of visualizations in motion, with an emphasis on user experience.

\begin{figure*}[tb]
    \centering
    \includegraphics[width=0.95\linewidth]{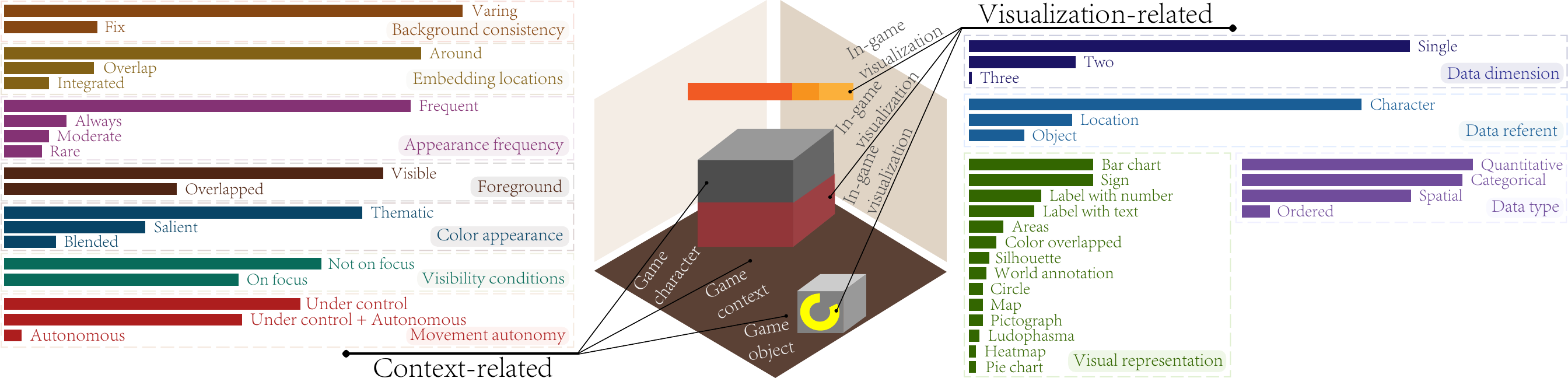}
    \caption{%
    An overview of the dimensions analyzed in our systematic review on in-game visualizations in motion. }
    \label{fig:teaser}
\end{figure*}

\section{Related Work}
Our research is closely related to work on situated and embedded visualization \cite{Willett:2017:EDR}.  
We focus on embedded visualizations that move with their referents as a specific type of visualization in motion \cite{Yao:2022:VisinMotion}.  
Since we select video games as our scenario, we additionally compare ours to previous work on visualizations in video games.

\subsection{Situated and embedded visualization}
Situated visualizations are data representations shown in close proximity to relevant referents such as objects, locations, or people. Bressa et al.\ provided an overview and survey of definitions and related work \cite{bressa:2022:WhatSituationSituated}. Willett et al. \cite{Willett:2017:EDR} formalized the difference between situated and embedded visualizations through the relationship among data and data referents. 
Many dedicated applications of situated and embedded visualizations were developed by using technology such as AR \cite{Chen:2021:VisCommentator, Chen:2023:iBall, Lin:2022:Omnioculars, Lin:2020:SportsXR, Lin:2021:ARVisforBasketballTraining}, mobile devices \cite{Grioui:2021:HeartRateVirtualSmartwatch, Amini:2017:DRF, Schiewe:2020:Smartwatch-Sports-Activities-Vis, Schildbach:2010:ReadingWhileWalking}, data physicalizations \cite{Taher:2017:ExplorationofDynamicPhysicalBar, Rodighiero:2018:PWV, Rodighiero:2021:MA}, or regular displays \cite{WU:2021:VRTableTennis, Wang:2021:TableTenis, SoccerTeamSportAnalysis:2017, Bai:2016:BasketballAnalysis}. 
This prior work targeted embedded visualizations for application scenarios by providing specific techniques and evaluations. We complement this work with dedicated visualization design guidelines for embedded \emph{visualizations in motion}.

Shin \etal\ conducted a survey on situated analytics \cite{Shin:2024:SituatedReality} by investigating 47 situated analytics systems. They classified the collected systems according to the situatedness of the visualizations, the data depictions, and the sensemaking support. They proposed several design guidelines for a better situated visualization design. For example, the authors suggested to add more sensemaking support, richer interactions and data depictions, and context awareness.  
Lee \etal\ \cite{Lee:2024:DesignSpace, Lee:2022:DesignSpace} systematically explored design patterns of situated visualization in augmented reality. Specifically, they investigated how the visualizations were situated, where the visualizations were placed, and the spatial and view coordinate relationships between the visualizations, their referents, and their viewers. They proposed design guidelines on how to place visualizations depending on different aspects of the real world, such as the location and the density of referents and the navigation between referents.
Moreover, Saffo \etal\ \cite{Saffo:2024:ImmersiveAnalytics} conducted a broader systematic review of 73 immersive analytic systems to explore features that affect system designs. Their review was on immersive analytics but also included some dimensions related to visual representation. They discussed data dimensions, representation types, visualization size, position, manipulations, and interactions. The authors further reflected on the importance of visual representation for designing immersive analytic systems and its influence on immersive technologies.
These previous reviews and proposed design spaces all discussed the position or the placement of visualizations in a broad way. In contrast, concrete research regarding different embedding locations of visualization and their impact on user experience has not yet been investigated. 

\emph{Embedding location} is one of the important criteria in many situated visualization definitions. The way we consider embedding location in our work is tightly related to what White and Feiner \cite{White:2009:SiteLensSV} call locus of presentation and spatial relevance in their situated visualization framework. Willett \etal\ \cite{Willett:2017:EDR} 
explain the difference between situated and embedded visualization.
In situated data representation, we find a spatial proximity relationship between data and its referent. 
Embedded representation takes this proximity to the extreme by tightening the data representation with its referent. Our designs use these tight spatial connections, and thus, we consider them embedded representations.

\subsection{Visualization in motion}
Yao \etal\ describe visualization in motion as visual representations used in contexts that exhibit relative movement between the viewer and the visualization \cite{Yao:2022:VisinMotion}.
Previous research on visualizations in motion focused on visualization perception and the impact of the motion context. Yao \etal\ conducted a series of perception studies that showed visualizations could be read under high speed and irregular trajectories \cite{Yao:2022:VisinMotion}. The authors evaluated both basic horizontal and circular bars (donuts) out of context. In our current work, we use their designs of horizontal and circular bars.
Related studies on visualization in motion come from psychology. Dynamic Visual Acuity (DVA) is the ability to visually discriminate details of objects under motion \cite{Quevedo:2018:DVA}. Research showed that DVA depends on the contrast between the moving target and the background against which it moves. Quevedo \etal\ \cite{Quevedo:2010:AssessDVA} highlighted that speed, trajectory, stimulus exposure time, and ambient illumination influence dynamic spatial resolution significantly. 
In contrast to exploring the visualization performance in a concrete scenario, these studies were done without context, and the reading task was the only primary task the participants were supposed to complete.

Follow-up work from Yao \etal\ \cite{Yao:2022:VisForSwimming} applied visualizations in motion in a sports scenario. The authors developed a technology probe called \emph{SwimFlow} to support users in designing and embedding visualizations in motion in sports videos. The focus of this work was the challenges of the design \emph{process} rather than on guidelines for visualization in motion design.
Other work on visualization in motion mainly focused on the concept and research challenges in different scenarios, such as exercising with fitness trackers \cite{Yao:2022:FitnessTrackers}.

To complement prior work, we conduct a contextual exploration of visualization in motion with a focus on the design considerations and trade-offs regarding user experience.    

\subsection{Visualizations in video games}
Video games are an application scenario in which visualizations are widely embedded to give players an immersive experience and to support players' tasks. These in-game visualizations have been investigated from multiple perspectives. For example, Zammitto \etal\ \cite{VisInGame:2008} analyzed how video games provide the player with important visual information through techniques such as silhouettes, mini-maps, Head-Up displays (HUDs), or Fog of war. The authors stressed the importance of reviewing visualizations by genre. We took this advice into account in our systematic review. Bowman \etal\ \cite{TowardVisualizationforGames} presented a framework of five categories, including primary purpose, target audience, temporal usage, visual complexity, and integration to classify any in-game visualization. 
Following these classifications, the visualizations designed in our study have the following characteristics: 
they provide status information for the player, are continuously present, have basic visual complexity to be read at a glance, and follow an integrated (situated) design.

Other research has investigated specific data visualizations and their effectiveness. A wide exploration was conducted on game characters' health data. Brooksby \cite{Brooksby:2008:RepresentationOfHealth} identified five categories of health representations in his exploratory study: mobility, ability, psychology, social, and pain (health), in which the most frequent depiction of health appeared to be pain. 
Peacocke \etal\ \cite{Peacocke:2018:EmpiricalComparisonFPS} studied players' performances with various types of displays. They used three health visualizations: a bar, heart icons forming a bar, and blood splatter. All three were compared as a HUD display and a diegetic \footnote{Game elements are diegetic if they exist within and are consistent with the fiction of the game’s world (e.\,g.\ the player’s character perceives them, and the player responds to them as such). Elements are not diegetic if they exist for the player but not for the characters (e.\,g.\ most of the HUD or third-person perspective information) \cite{Peacocke:2018:EmpiricalComparisonFPS, Marre:2021:DiegeticInterfaces}.}
design. Diegetic designs are most closely related to situated visualizations. Their results showed that participants favored the HUD variants and that around half of the players found the diegetic designs to be a hindrance to their game success.
Dedicated research on static visual representation preference existed.
Gittens and Gloumeau \cite{Gittens:2015:SegmentedHealthBar} explored the impact of segmented health bars with a fixed position on the screen on players' preferences for a game. Their results showed that most of their players preferred segmented bars over a single-bar design. Moving visualizations were not explored, indicating their designs might have challenges.

Other work focused on the gaming experience and games' playability \cite{Hoobler:2004:VisCompetitiveBehaviour, Clarke:2006:HowGamersExperienceGS, El-Nasr:2006:VisualAttention, Fabricatore:2002:PlayabilityDesignModel, Martinez:2012:StudyOfColorInGameEngagement, Karlsson:2016:InformationVI, Dillman:2018:VisualInteraction, Horbinski:2021:MapSymbols}. El-Nasr  \etal\ \cite{El-Nasr:2006:VisualAttention} showed that players focus on the center of the screen in FPSs but look around more widely in adventure games. Dillman \etal\ \cite{Dillman:2018:VisualInteraction} investigated how visual interaction cues like an augmented arrow can guide participants where to look and go in a game. Horbi{\'n}ski and Zagata \cite{Horbinski:2021:MapSymbols} studied visual symbols on a map in the survival game \emph{Valheim}. The researchers used pictographs to show information to players. Martinez \etal\ \cite{Martinez:2012:StudyOfColorInGameEngagement} studied color and found a correlation between game type and the color used. Also, Karlsson \etal\ \cite{Karlsson:2016:InformationVI} stated how important colors are to distinguish enemies from allies in FPS games and Massively Multiplayer Online Role Playing Games (MMORPG). 
We decided to use bold colors such as turquoise, light green, and yellow for the ambient light of our video game to convey excitement and allow for a good contrast between the visualization and the ambient colors.
Motion has also received attention in the context of user experience (but not for visualizations). Milam \etal\ investigated motion attributes' effects \cite{Milam:2011:cameramotiongame}, game element designs \cite{Milam:2012:designpatterns}, and visual features \cite{Milam:2012:designsSimilarity}. Their work concentrated on the impact of game element designs on gaming experience or the effects of gaming expertise on game performance but not on visualizations. 
What has not been researched yet is the impact of \emph{visualizations in motion} designs on game experience. We specifically focus on this challenge.

\begin{table}[tb]
    \centering
    \captionsetup{justification=centering}
    \caption{Representations of collected in-game visualizations in motion}
    \footnotesize
    \renewcommand{\maxnum}{36}
    \tabulinesep=0mm
    \aboverulesep = 0.5mm
    \belowrulesep = 0.5mm
    \begin{tabu}{
     @{\hspace{0ex}}R{2.3cm}@{\hspace{1ex}}C{0.8cm}@{\hspace{1ex}}C{0.9cm}@{\hspace{1ex}}
     | @{\hspace{1ex}}R{2.3cm}@{\hspace{1ex}}C{0.8cm}@{\hspace{1ex}}C{0.9cm}@{\hspace{0ex}}}
        \toprule
            \textbf{Representation} & \textbf{Sample} & \textbf{Counts} & \textbf{Representation} & \textbf{Sample} & \textbf{Counts}\\
        \midrule
        
        Sign & \includegraphics[height=2.5ex]{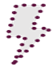} & \databar[isoorange]{36} & World annotation & \includegraphics[width=4ex]{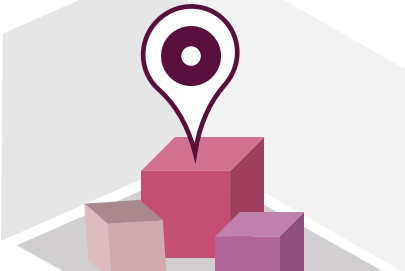} & \databar[isoorange]{5}\\
        \midrule

        Bar chart & \includegraphics[width=6ex]{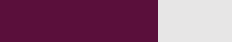} & \databar[isoorange]{36} & Circle & \includegraphics[height=2.5ex]{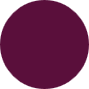} & \databar[isoorange]{4}\\
        \midrule

        Label with number & \includegraphics[height=2.5ex]{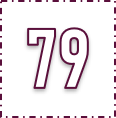} & \databar[isoorange]{21} & Map & \includegraphics[height=2.5ex]{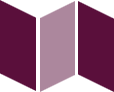} & \databar[isoorange]{4}\\
        \midrule
        
        Label with text & \includegraphics[height=2.5ex]{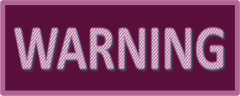} & \databar[isoorange]{19} & Pictograph & \includegraphics[height=2.5ex]{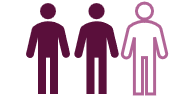} & \databar[isoorange]{4}\\
        \midrule

        Areas & \includegraphics[height=2.5ex]{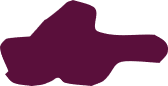} & \databar[isoorange]{10} & Ludophasma & \includegraphics[height=2.5ex]{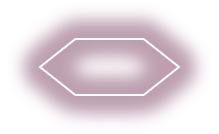} & \databar[isoorange]{3}\\
        \midrule

        Color overlapped & \includegraphics[width=6ex]{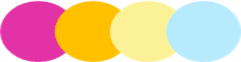} & \databar[isoorange]{8} & Heatmap & \includegraphics[height=2.5ex]{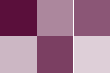} & \databar[isoorange]{2}\\
        \midrule

        Silhouette & \includegraphics[width=5ex]{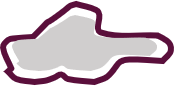} & \databar[isoorange]{6} & Pie chart & \includegraphics[height=2.5ex]{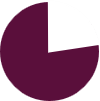} & \databar[isoorange]{2}\\
        \bottomrule
    \end{tabu} 
    \label{table:visualRep}
\end{table}

\section{In-game Visualizations in Motion:\\ A Systematic Review}
Visualizations in motion are prevalent in video games to guide players through tasks. Their designs need to be adapted to fit the motion context while allowing people to focus on the game and get helpful contextual information at the same time. We consider existing in-game visualizations in motion as artifacts that can be studied to understand decisions designers made to adapt data representations to this context.

To understand how designers dealt with the various challenges related to motion in video games, we reviewed visualizations from 50 games and 17 different genres. In total, we found 160 examples of visualizations in motion that we categorized according to different dimensions (\autoref{fig:teaser}). 
In the development of our categories, we were inspired by the work of Islam \etal\ \cite{Islam:2020:VisOnWatchFaces} in their analysis of existing visualizations in use on watch faces. The dimension used by Islam \etal\ described the visual design of data representations and the data used, which we also included. We extended their categorization by adding dedicated dimensions related to motion factors and video games. To further understand how different data are shown, we also classify the representations designers chose for different data types. A small portion of our systematic review was published in the format of posters \cite{Bucchieri:2022:EuroVis, Bucchieri:2022:DataType}.
Next, we describe our game selection, visualization categorization, and the results that emerged from our systematic review.

\subsection{Selection of Video Games}
To cover a diverse selection of video games we used a commercial ranking website called Metacritic \cite{Metacritic}. 
Metacritic assigns a unique Metascore to each video game and categorizes games according to 18 different game genres. For each genre, we selected the top 3 games from 2011---2022 across all gaming platforms.
We excluded puzzle games because, on first inspection, they did not contain moving visualizations. Moreover, for the wrestling genre, we found and selected only two games that contained relevant visualizations. We excluded games without a Metascore and maintained one of the games from the same series (\eg, \textit{Super Mario Galaxy} \cite{supperMarioGalexy} and \textit{Super Mario Galaxy 2} \cite{supperMarioGalexy2}). In total, we reviewed 50 games from 17 genres (Appendix-\autoref{table:selectedGames}). For each game, we watched game-plays on YouTube for approximately 5 to 15 minutes and video-recorded relevant parts of the videos where the game showed visualizations in motion.

\begin{table}[tb]
    \centering
    \footnotesize
    \captionsetup{justification=centering}
    \caption{Embedding locations and movement autonomy of collected data types visualized under motion}
    \renewcommand{\maxnum}{68}
    \newcolumntype{H}{>{\centering\arraybackslash} m{1.4cm}}
    \tabulinesep=0mm
    \aboverulesep = 0.5mm
    \belowrulesep = 0.5mm
    \begin{tabu}{ 
     @{\hspace{0ex}}H 
     |@{\hspace{0ex}}C{1.1cm}@{\hspace{0ex}}
     |@{\hspace{0ex}}C{1.1cm}@{\hspace{0ex}}
     |@{\hspace{0ex}}C{1.1cm}@{\hspace{0ex}}   
     |@{\hspace{0ex}}C{1.3cm}@{\hspace{0ex}}
     |@{\hspace{0ex}}C{1.2cm}@{\hspace{0ex}}
     |@{\hspace{0ex}}C{1.1cm}@{\hspace{0ex}}
    }
        \toprule
        \multirow{2}{*}{\shortstack{\textbf{Data type} \\ \textbf{\& Counts}}} & \multicolumn{3}{c|}{\textbf{Embedding locations}} & \multicolumn{3}{c}{\textbf{Movement Autonomy}}\\
        \cline{2-7}
        & \small{Around} & \small{Overlap} & \small{} & \small{Autonomous} & \small{Controlled} & \small{Mixed} \\
        \midrule
        \shortstack{Quantitative\\\databar[isoorange]{68}} & \databar[isoblue]{56} & \databar[isoblue]{6} & \databar[isoblue]{6} & \databar[isogreen]{2} & \databar[isogreen]   {37} & \databar[isogreen]{29}\\
        \midrule
        \shortstack{Categorical\\\databar[isoorange]{64}} & \databar[isoblue]{49} & \databar[isoblue]{10} & \databar[isoblue]{5} & \databar[isogreen]{2} & \databar[isogreen]{36} & \databar[isogreen]{26}\\
        \midrule
        \shortstack{Spatial\\\databar[isoorange]{49}} & \databar[isoblue]{29} & \databar[isoblue]{12} & \databar[isoblue]{8} & \databar[isogreen]{0} & \databar[isogreen]{32} & \databar[isogreen]{17}\\
        \midrule
        \shortstack{Ordered\\\databar[isoorange]{8}} & \databar[isoblue]{8} & \databar[isoblue]{0} & \databar[isoblue]{0} & \databar[isogreen]{1} & \databar[isogreen]{0} & \databar[isogreen]{7}\\
        \bottomrule
    \end{tabu}
    \label{tab:AnalysisOfDataTypes}
\end{table}

\begin{figure}[tb]
    \centering
    \includegraphics[width=\linewidth]{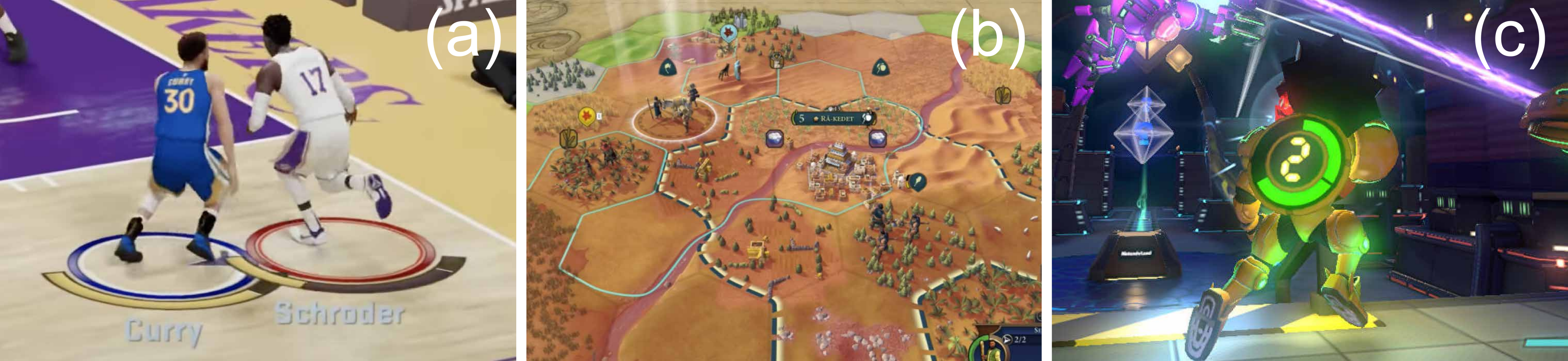}
    \caption{Examples of visualizations in motion embedded in different locations with respect to the data referents: (a) Stamina bars and names embedded under the feet of the players in \emph{NBA 2K21} \cite{NBA_2K21}; (b) Heatmap overlapping the game environment in \emph{Civilization VI} \cite{Civilization6}; (c) Health counter  with the character's suite design in \emph{Nintendo Land} \cite{nintendo_land}.}
    \label{fig:embeddinglocations}
    \vspace{-5mm}
\end{figure}

\begin{figure*}[tb]
    \centering
    \includegraphics[width=.95\linewidth]{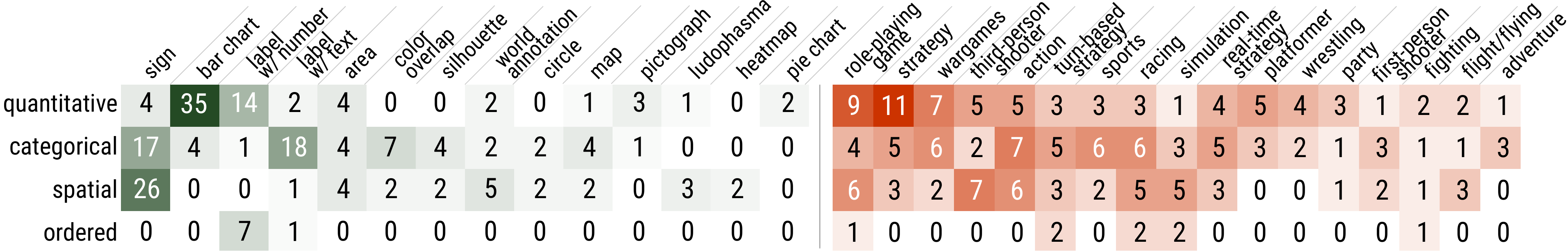} 
    \caption{Distributions of collected data visualized under motion per visual representation (left) and per game genre (right).}
    \label{fig:dataTypeDistribution}
\end{figure*}

\subsection{Categorization of Current In-game Visualizations}
Among our 160 collected in-game visualizations in motion, the genres with the most occurrences were Role-playing Games (RPG), with 18/160 visualizations; strategy games, with 16/160; and wargames, with 14/160 samples. We categorized these visualizations (process described in Appendix-Section C.2) based on multiple dimensions related to situated visualization and motion characteristics:  
\vspace{1pt}
\noindent\textbf{Visual representation:} describes how data was represented. Signs (\eg, icons, arrows) and bar charts--both linear and circular--were the most prevalent representations (\percentagebar{0.225} 36/160), followed by labels with numbers (\percentagebar{0.13} 21/160) and labels with texts (\percentagebar{0.12} 19/160). \autoref{table:visualRep} summarizes all the representations found. 

\vspace{1pt}
\noindent\textbf{Data referents:} are the entities the data refers to. We found three types of referents: characters, locations, and objects. A character is an actor in the game controlled by a player or an AI together with its own visible equipment like guns, armors, or clothes. A location is an area or point of interest inside the scope of the game (\eg, objective's position, area of attack, checkpoints). An object is any nonliving entity that the player can interact with. Most \percentagebar{0.72} 114/160 data referents were characters,  followed by \percentagebar{0.18} 30/160 locations, and \percentagebar{0.10} 16/160 objects. 

\vspace{1pt}
\noindent\textbf{Data dimensions:} refer to the data represented in each visualization. Visualizations showing only a single dimension were the most common case (\percentagebar{0.80} 128/160), while 31/160 samples encoded two dimensions and only one encoded three types of information. Often, additional dimensions were derived from a primary dimension. For example, in addition to showing a character's health percentage, various games applied color encoding to represent a health status. A widely used encoding used a traffic light color scheme with green, yellow, and red for healthy, medium, and critical health levels. 

\vspace{1pt}
\noindent\textbf{Color appearance:} refers to how the visualizations' colors were integrated. We divided color appearance into three groups: blended, salient, or thematic. \emph{Blended} colors integrate the visualization with the game context. The visualization does not visually stand out. This may lead to visualizations being overlooked or difficult to read but may create a more immersive experience. \emph{Salient} colors make the visualization highly visible; either through the use of untypical hues or through a very high contrast with the game context. \emph{Thematic} colors are either common for the data they represent or match the video game’s palette. The color appearances of \percentagebar{0.65} 104/160 samples were thematic, while 41/160 were salient, and only 15/160 were blended. 

\vspace{1pt}
\noindent\textbf{Background consistency:} considers if the visualization has a static or fixed background when a visualization changes its in-game position. When enemies move from one location in the game to another, the visualizations attached to them may be displayed on changing backgrounds. If the background of a visualization is fixed, for example, by providing an opaque container behind it, there is a higher possibility that the visualization will always be visible regardless of motion factors. Results show that \percentagebar{0.83} 133/160  visualization's backgrounds were varying with the visualization position, while only 27/160 had a fixed background. 

\vspace{1pt}
\noindent\textbf{Embedding locations:} indicate the spatial relationship of visualization and data referent. We found three different embedding locations. \percentagebar{0.75} 121/160 visualizations were embedded \emph{around the data referent}; for example above an object (\autoref{fig:embeddinglocations}(a)), under the feet of a character, or above a checkpoint (a location). 26/160 visualizations showed full or partial \emph{overlap with the data referent} as shown in \autoref{fig:embeddinglocations}(b). Finally, 13/160 visualizations were \emph{integrated into the data referent} permanently, for example, in its material color, in its design, or in the diegetic elements attached to it (\autoref{fig:embeddinglocations}(c)). Although visualizations integrated or overlapped with the referent might look similar, integrated visualizations cannot be separated from the referent while overlays may be temporary. 

\vspace{1pt}
\noindent\textbf{Movement autonomy:} considers if the visualization was moving \emph{autonomously} or whether movement was \emph{caused by a player}. For example, in a third-person shooter (TPS), the player controls a main character defined as the protagonist. By moving the protagonist, rotating its view, and changing the camera perspective, the player can induce the motion of elements on the screen. Autonomous movement, instead, does not depend on the player's control, for example, when the camera is still and some characters are moving on screen. Autonomous movement was not prevalent in video games. Only 5 visualizations in our collection moved autonomously, while \percentagebar{0.53} 86/160 were consistently controlled in some way by the player and \percentagebar{0.43} 69/160 depended on the mix of autonomous movement and the player's control.

\vspace{1pt}
\noindent\textbf{Visibility conditions:} considers if the visualization is only visible based on player interaction.  Examples include hovering over the reference, pushing a button, or using a special power. We consider those visualizations to be \textit{on focus}. If they are visible independently of the player's actions, they are \textit{not on focus}. ``On focus'' is very common in FPS games where enemy health bars are only visible while the cross-hair of the protagonist hovers over the body of the enemies themselves. The slight majority of the cases were ``not on focus'' (\percentagebar{0.57} 92/160), while \percentagebar{0.43} 68/160 visualizations were ``on focus.''

\noindent\textbf{Appearance frequency:} evaluates how often a visualization in motion appears during a normal gaming session. Frequency is a relative measure which we split into four categories: Rare (11/160) indicates occasional appearance (\eg, progress bars of special actions); Moderate (13/160) represents visualizations that appeared from time to time (\eg, secondary tasks related visualizations such as troops movement in resource management games); Frequent (\percentagebar{0.73} 118/160) defines visualizations appeared often but not continuously (\eg,  primary task-related visualizations like health bars on focus for FPS games); Always Visible (18/160) means the visualizations were continuously displayed on the screen (\eg, the health of the main character in TPS game). 

\vspace{1pt}
\noindent\textbf{Foreground:} indicates if the visualization is always in the foreground and cannot be overlapped. Occlusion is an important factor to consider in video games. If we imagine a scene with 20 enemies in front of the game protagonist, each one of them with a visualization displayed under their feet, it is likely that some of the visualizations will be poorly visible or hidden due to the occlusion caused by other elements. For example, a tree can hide some visualizations displayed nearby. Being always in the foreground can help the visualizations to stand out. 110/160 \percentagebar{0.69} visualizations were always in the foreground, while  50/160 \percentagebar{0.313} visualizations could be overlapped by other game elements.

\subsection{Representations by Data Type}
The four most prevalent data types we found in our review were quantitative, categorical, ordered, and spatial data. 
A single visualization can represent multiple data types depending on the encoding used. For instance, a health bar can encode both quantitative (health points) and categorical (team affiliation) data. Similarly, a game mission icon can encode spatial (distance) and categorical (target type) data.
We also analyzed the spatial relationship of the visualization and the data referent. There were some tendencies in how different types of data were \emph{embedded} in the data referent. The vast majority of each data type was embedded around the data referent, with one exception highlighted below. Regarding the \emph{movement autonomy} of different types of data representations, most visualizations were in motion due to the player's control. Details per data type we analyzed are represented in \autoref{tab:AnalysisOfDataTypes} and \autoref{fig:dataTypeDistribution}. Following, we highlight the most interesting findings: 

\vspace{1pt}
\noindent\textbf{Quantitative data:} was the most common data type, which was mainly represented by bar charts (both linear and circular ones) and labels with numbers.
Examples include the health points of a character, its stamina, or the number of resources crafted. Quantitative data was very popular in strategy games, RPGs, and wargames. 

\vspace{1pt}
\noindent\textbf{Categorical data:} was the second most common, concerned team identification and resource type. It was represented as labels with text, signs, or by color-overlap on the data referent. Categorical data was frequently embedded in action, racing, sports, and war games.

\vspace{1pt}
\noindent\textbf{Spatial data:} refers to position-related information. The position was most commonly indicated by signs. This data was very common for mission objectives or interactive item positions. Also, the use of world annotations was common for spatial data. 
Spatial data was often used in TPS games, action games, and RPGs.

\vspace{1pt}
\noindent\textbf{Ordered data:} was the least frequent data type. It was used in specific game genres like racing, simulation, and TPS games (\eg, for the position on a leaderboard). Ordered data was represented with labels using numbers or text and that \emph{only} around the data referent. Ordered data was also the only type of data that showed a majority of samples with movement autonomy depending on mixed of controlled motion and autonomous motion (\percentagebar{0.875} 7/8). 

\subsection{Summary}
We examined how designers addressed the challenge of designing in-game visualizations in motion by reviewing the produced artifacts. 
We found mostly small visualizations embedded around game characters, conveying a single piece of information like health; potentially together with the display of a derived value (\eg, an indicator to show how critical a health value may be).
Designers used well-known representations such as signs (icons or arrows), length encodings (round, horizontal, or vertical bar charts), or text. 
Designers used different strategies to integrate the visualizations in motion into the game:  visualizations were often placed around  a character with an adjusting background that made the visualizations distinguishable from the rest of the scene. Visualizations predominantly used color palettes that fit the game theme or were canonical colors. As a result, visualizations were not blended deeply into the game world but neither stood out excessively. 
We found this latter finding particularly intriguing, as a deeper embedding could potentially improve the feeling of immersion and user experience but affect performance. Thus, we aimed to study more deeply how different in-game visualization in motion designs affect user experience. 

\section{RobotLife: A Testbed for Evaluation}
\begin{figure}[t]
    \centering
    \includegraphics[width=\columnwidth]{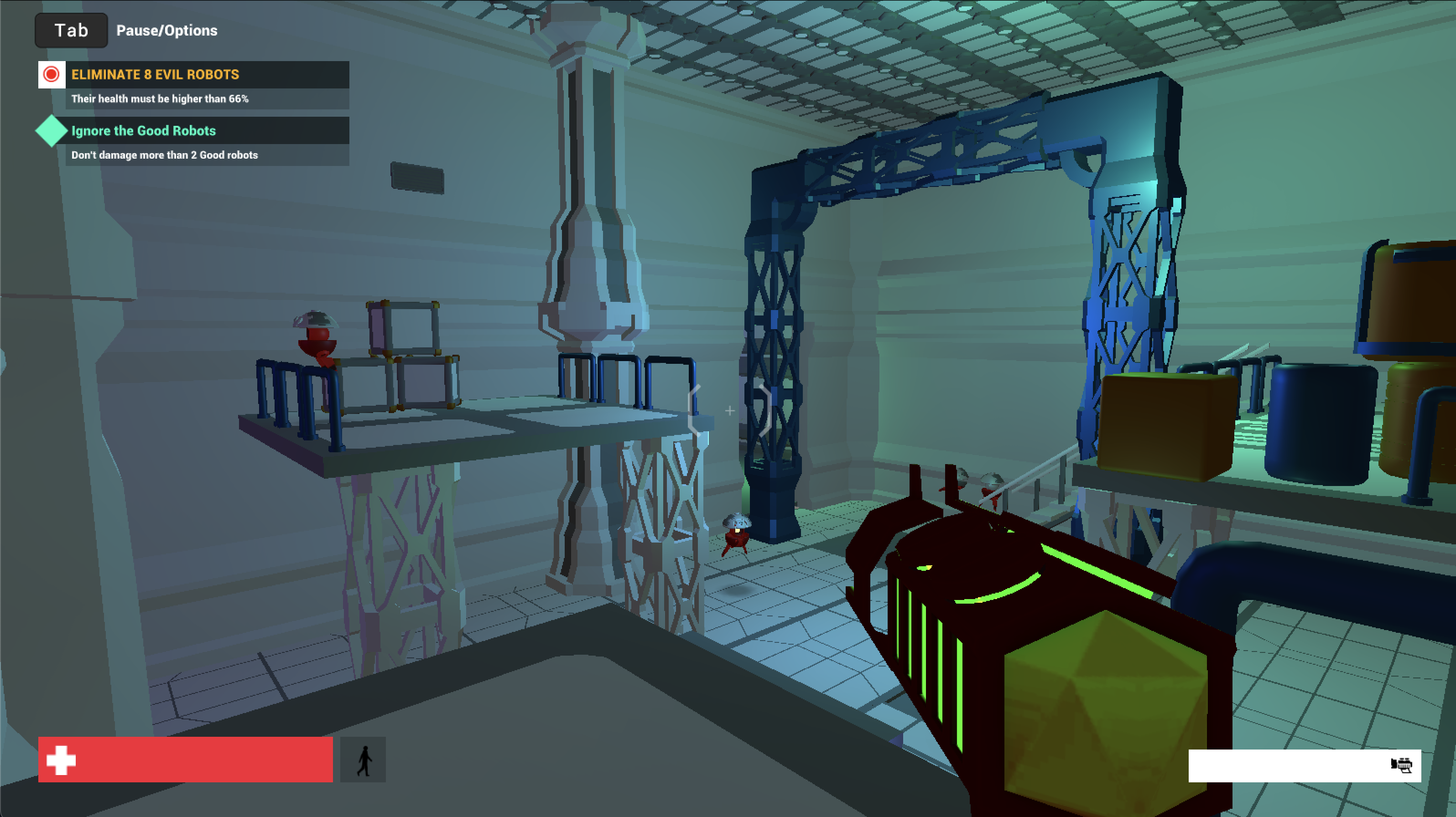}
    \caption{A screenshot of \emph{RobotLife}'s UI. The player's health and ammunition \revisionVIS{count} are shown at the bottom in red and white bar charts respectively. Primary and secondary objectives of the game are displayed on the left of the screen, together with a pause button.}
    \label{fig:UIscreenshot}
\end{figure}

To be able to customize game tasks, collect game data, and embed visualizations in motion, we developed our own testbed platform rather than a representative game design: the  FPS game \emph{RobotLife} (\autoref{fig:UIscreenshot}). While FPS was not the genre that included the most visualizations,\footnote{We note that a single video game can fall under different categories, and many games in other Metacritic categories have close relationships to FPSs.} a FPS game allowed us to include several visualizations with quick autonomous movement (associated with enemies) and movement controlled by the player. Moreover, FPS games are popular \cite{Konnikova:2013:FPSPupularity} which would allow us to find experienced players for our user study. 

\subsection{Storyline}
\emph{RobotLife} is set inside a robot factory. A virus infiltrated the factory and altered some robots. The altered robots became evil and started damaging the electric system of the factory. The player acts as the guardian of the factory and has the duty to eliminate all evil robots. This task comes with a main challenge: identifying the evil robots. The evil robots act secretly and try not to be recognized while destroying the factory; yet, the virus gave extra health to the evil robots, so they can be identified by looking at their health indicators: good robots' health was $<$66\% and evil robots' was $>$66\%.
We made this decision because, in order to understand challenges in visualization reading, we wanted players to pay attention to the visualization instead of acting by inertia.
Robots are not dangerous to humans, but every robot comes with a life-saving mechanism, so when attacked, they fire back immediately. The central management system of the factory spotted 8 evil robots connected to the electrical system of the facility. The player wins the game when all 8 evil robots are eliminated and not more than 2 good robots were damaged. An evil robot can never become a good one. Robots move in the entire factory under irregular trajectories, in different directions, and at changeable speeds. The level map was designed to be linear but at the same time challenging and big enough to support navigation without information overload. Some areas of the map encourage players to jump between platforms or over decorative objects, with the aim of creating extra motion factors in the gameplay.

\subsection{Implementation}
\begin{figure*}[t]
    \centering
    \includegraphics[width=\linewidth]{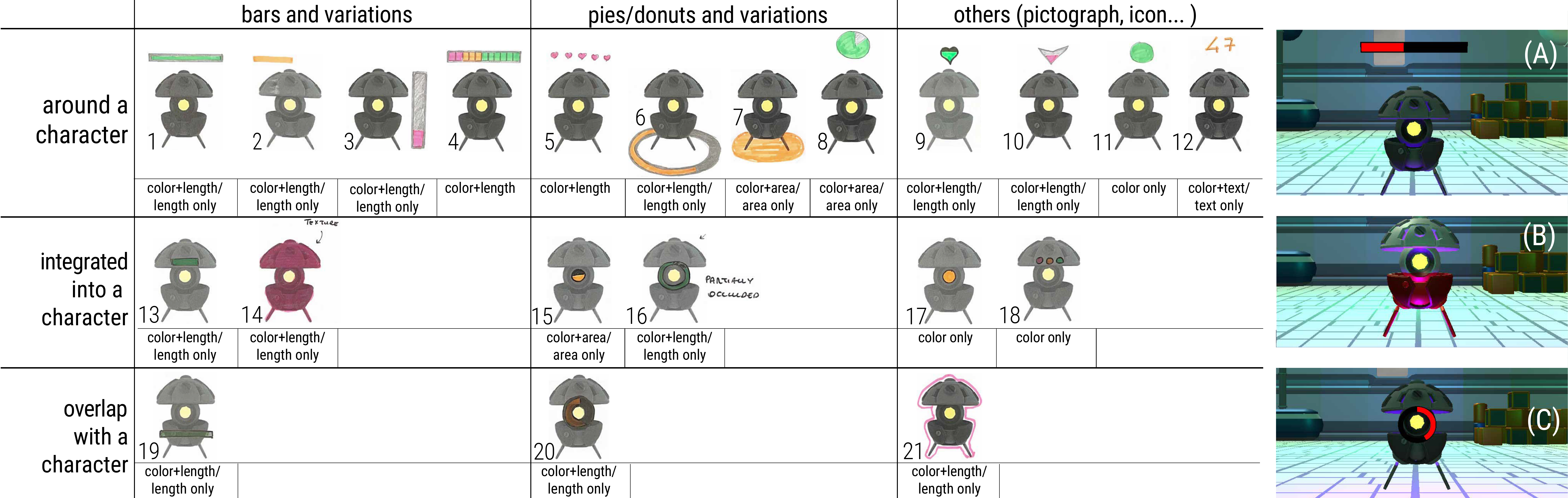}
    \caption{Left: Health visualizations considered using different embedding locations, colors, and types of encoding. Other design variations are easily possible for Rows 2 and 3 by using ideas from Row 1. These were not explicitly drawn out. Right: Three visualization designs illustrated in \emph{RobotLife}. (a) \includegraphics[height = 0.32cm, valign=c]{Figures/icons/bar.pdf}\ \texttt{Non-integrated design}: a horizontal bar chart positioned outside of the robot; (b) \includegraphics[height = 0.32cm, valign=c]{Figures/icons/texture.pdf}\ \texttt{Fully-integrated design}: a vertical bar chart integrated into the texture of the robot; and (c) \includegraphics[height = 0.32cm, valign=c]{Figures/icons/donut.pdf}\ \texttt{Partial-match design}: a circular bar (donut) matching a part of the robot's shape.}
    \label{fig:sketches}
    \vspace{-6mm}
\end{figure*}

We implemented \emph{RobotLife} based on Unity, a cross-platform game engine \cite{Unity} and by extending an \emph{FPS Microgame Template} \cite{FPSMicrogame:2022:Unity}. 
The FPS Microgame Template is a free-available 3D first-person shooter game on the Unity Learn platform and can be modified and customized by developers. The microgame comprises level-building assets, weapons, props, enemies, and more. The microgame comes with basic game mechanics: a first-person character controller for recognizing player inputs, a rudimentary environment in the sample scene, and shooting mechanics to combat enemies within the scenario. The microgame also proposes designs and implementation for hover bots and the flying robots inside the factory (\autoref{fig:sketches}: Right). 

FPS games are often centered around killing all the enemies without distinction. Players do not need to read the exact value of an enemy's health representation, as their final goal is always to bring the health value to 0. For this reason, the mechanics of our game \emph{RobotLife} were developed in a different way, forcing the players to actually read the health level to complete their gaming tasks and not blindly shoot at every enemy in the scenario.
We controlled the autonomous movement of hover bots (and their visualizations) to ensure that the relative motion factors (\eg relative speed, direction) between these visualizations and the player were of adequate magnitude to provide a challenge in game play and for reading the visualizations in motion. 

\section{An Evaluation of Visualizations in Motion:\\ Study Design}
Our main goal is to explore how players would experience visualizations in motion in a busy context that required a primary task: primarily navigation or shooting. 
To understand the player experience, strictly-controlled experiments with simple and short tasks are not the right choice, particularly in video games. FPS gameplay in practice is a continuous process with dynamic changes in backgrounds. Players need to react quickly to a situation that continuously evolves depending on their actions. Players' gaming strategies cannot be strictly controlled to be the same for each player. 
Thus, we designed our study around a realistic gaming experience. We analyzed both qualitative and quantitative indicators of game performance and user experience. Next, we present the in-game visualizations in motion we designed, our study task, procedure, pilot, and participants.

\subsection{Visualization Design and Representation Selection}
In our systematic review, in-game visualizations in motion most often depicted quantitative data and single-dimensional health values. As such, we implemented visualizations of health data for each robot.

We chose the \emph{integration} level of visualization to its data referent as the main varying factor. The integration level is composed of the visualization's embedding location and its representation, which cannot be discussed individually. On the one hand, the embedding location (\ie, around a game character,  with the character, and overlapping a game character) can have a direct impact on how viewers see and experience the visualizations themselves in terms of aesthetics, immersion, and readability. On the other hand, the embedding location has an impact on the selection and sensation of encoding types.
For example, a visualization displayed around the character highlights its position and visibility but lets the visualization artificially stand out; a visualization integrated with the character's design impacts the character's aesthetics by altering its original design and may be less readable; while a visualization overlapped with the character is more prone to occlusion due to other objects in the game scenario. 

\autoref{fig:sketches}-Left illustrates an initial set of visualizations that we designed inspired by our systematic review, presenting different combinations of embedding locations, visual representations, colors, and encoding types. More design variations are achievable by choosing different colors, adjusting spatial arrangements, and making changes. Instead of constructing nearly endless possibilities, we focused on these 21 options as they already offer a good diversity of integration levels. 
According to our pilot test, we found that a reasonable completion time for our formal study allowed us to test three visualizations per participant. To determine reasonable visualizations, we prioritized the designs that we saw in actual video games as well as readability in our game context. We also chose designs that required the same type of visual inference: the reading of length, which was one of the most common encodings in our systematic review. We excluded Designs 8, 12, 19, and 21 as they were not implemented similarly in any actual games. We could not use Designs 6 and 7 embedded below the character because the robots floated, and the distance from the ground could be confusing for players. We also excluded Design 16 due to the partial occlusion caused by the robot's design. Designs 10, 13, 15, 17, and 18 are too small to be quickly readable while 9 and 11 as primarily categorical encodings.
We finally chose our designs to be based on bar charts, which were the most prevalent representation according to our systematic review, and some bar designs were already evaluated as readable in other non-contextualized perception study of visualization in motion \cite{Yao:2022:VisinMotion}. We chose Design 1, a typical horizontal bar over the head of the game character, as a  \barIcon\ \texttt{non-integrated design}. Design 3 was a close contender, but we chose the horizontal design as it was the most typical design in FPS and had been evaluated before \cite{Yao:2022:VisinMotion}. We excluded Designs 2, 4, and 5 as variations of Design 1 that have not been evaluated in visualization in motion perception. We next chose Design 14, a vertical bar that entirely into the character as a  \textureIcon\ \texttt{fully-integrated design}. And Design 20, a circular bar (donut) that overlapped on the game character as a \donutIcon\ \texttt{partial-match design}, which was also evaluated out of context \cite{Yao:2022:VisinMotion}. Design 16 was a circular bar with a thinner donut ring. In contrast, we followed Yao \etal's work to set the length and width of the horizontal bar (\barS) equal to the perimeter and ring thinness of the circular bar (\donutS).
Our three designs' in-game implementation is in \autoref{fig:sketches}-Right.

\subsection{Main Task and Study Variables}
In \emph{RobotLife}, players had to identify and eliminate 8 evil robots to win. Hitting more than 2 out of 16 good robots, regardless of killing them, resulted in losing. 66\% is a threshold for deciding if a robot is good or evil: good robots' health value was $<66\%$ health while evil robots had more. This rule prevented players from blindly shooting at robots. Each robot displayed one of 6 percentages: 18\%, 32\%, 43\%, 58\%, 72\%, and 83\% \cite{Yao:2022:VisinMotion}. To eliminate learning effects, each percentage value was randomly assigned to one of the 24 robots on the floor, giving each robot a different health value in every game iteration.

\subsection{Procedure and Measures}
The study took place in the lab, utilizing a DELL U3014t display (refresh rate: 60Hz), a Dell XPS 8910 desktop with an NVIDIA GeForce GTX 730 graphics card, a DELL KB216t keyboard, and a DELL MS116t1 mouse. The study was a within-subjects experiment where participants tested all three designs.
During the initialization phase, participants completed a pre-questionnaire regarding their demographics, gaming experience, and familiarity with PC gaming actions. Participants self-evaluated their proficiency in video gaming and PC gaming on a 7-point Likert scale. Additionally, they indicated the types of video games they usually played and reported their gaming frequency regarding games in general, FPS games, and PC games.

Participants proceeded to a tutorial to learn how to play the game and read visualization. They then completed three conditions, with the order determined by a Latin square. Each condition included both training and trial phases. 
In the training phase, we asked participants to do proportion estimation to determine whether a robot was good or evil in a static context in which the robot did not move. To advance to the trial phase, participants had to succeed at least 6 times but could do as many training trials as they wanted.
In the trial phase,  the gameplay consisted of playing \emph{RobotLife} for a duration of
5 minutes, with the possibility of multiple playthroughs. If a player died or won before the time elapsed, they restarted until the full duration was completed. 

Following the game, an in-game questionnaire evaluated the aesthetics of visualization designs outside their embedding context by using a 7-point Likert scale named BeauVis \cite{He:2023:BeauVis} and visualizations' readability outside the context through another 7-point Likert scale with 4 Likert items (\emph{clear}, \emph{readable}, \emph{intuitive}, and \emph{understandable}). After answering the in-game questionnaire, a new training-trial block started. Upon the player completing all three blocks, the study concluded with a post-questionnaire and brief post-interview.

In the post-questionnaire, participants ranked the three visualization designs by considering their embedding context: visualization designs' (a) readability during motion, (b) aesthetic fit with the game, (c) alignment with the game mission, and (d) contribution to a positive overall game experience. Participants explained the reasons for their ranking.
Finally, we interviewed participants, asking about (a) the impact of motion factors (both the movement of robots and players) on their gaming performance, (b) how contextual factors (\eg, background, lighting) affected their performance, (c) their strategies to read visualizations in motion, and (d) their general comments on our study.

\subsection{Pilot Study}
In order to thoroughly test our study, we conducted a pilot with 12 participants who were not regular game players (self-evaluated game-playing proficiency on video games: \emph{M} = 3.83/7, \emph{SD} = 1.34). Pilot details can be found in the related master thesis \cite{Bucchieri:2022:VisInGamesThesis}. 
We contrast the results of this pilot to our formal experiment when it is interesting. 

\subsection{Participants}
We looked for participants via university mailing lists and hung up flyers in university buildings. Following our institute's payment policies, participants received no remuneration in cash. Instead, we offered boxes of chocolate of equal value. In total, we recruited 18 game players:  2 \female, 16 \male. 
Participants reported their ages from 18---44 years: \ageRange, with most participants in the range of 25---34 years.
Our participants had high gaming expertise (\emph{M} = 5.80/7, \emph{SD} = 0.69) and were skilled with gaming using a keyboard and a mouse (\emph{M} = 5.17/7, \emph{SD} = 1.10).
The majority of our participants \percentagebar{0.833}15/18 usually played FPSs.
Overall, all our participants were video game enthusiasts. Over half of them had a high FPS gaming experience, and the majority of them frequently play video games on PC. Detailed information about our participants can be found in Appendix-\autoref{tab:gamefamilarity}.

\begin{table}[tb]
    \centering
    \captionsetup{justification=centering}
    \footnotesize
    \caption{Gaming performance scores across all 18 participants: Every single square represents a participant.} \includegraphics[height = 0.32cm, valign=c]{Figures/icons/bar.pdf}: \texttt{non-integrated design}, \\ \includegraphics[height = 0.32cm, valign=c]{Figures/icons/donut.pdf}: \texttt{partial-match design}, \includegraphics[height = 0.32cm, valign=c]{Figures/icons/texture.pdf}: \texttt{fully-integrated design}
    \tabulinesep=0mm
    \aboverulesep = 0.5mm
    \belowrulesep = 0.5mm
    \begin{tabu}{@{\hspace{0ex}}C{0.6cm}@{\hspace{0ex}}
                    |@{\hspace{0.5ex}}C{1.3cm}@{\hspace{0.5ex}}
                    |@{\hspace{0.5ex}}C{2.25cm}@{\hspace{0ex}}
                    |@{\hspace{0.5ex}}C{2.25cm}@{\hspace{0.5ex}}
                    |@{\hspace{0.5ex}}C{0.9cm}@{\hspace{0.5ex}}
                    |@{\hspace{0.5ex}}C{0.9cm}@{\hspace{0ex}}
                    }
        \toprule
        &
        Evil robots killed in total & 
        Distribution of how many players won how many times & 
        Distribution of how many players lost how many times & 
        Win\% means and CIs &
        Loss\% means and CIs \\
        \midrule
        
        \includegraphics[height = 0.6cm, valign=c]{Figures/icons/bar.pdf} &
        457 & 
        \includegraphics[width = 2.25cm, valign=c]{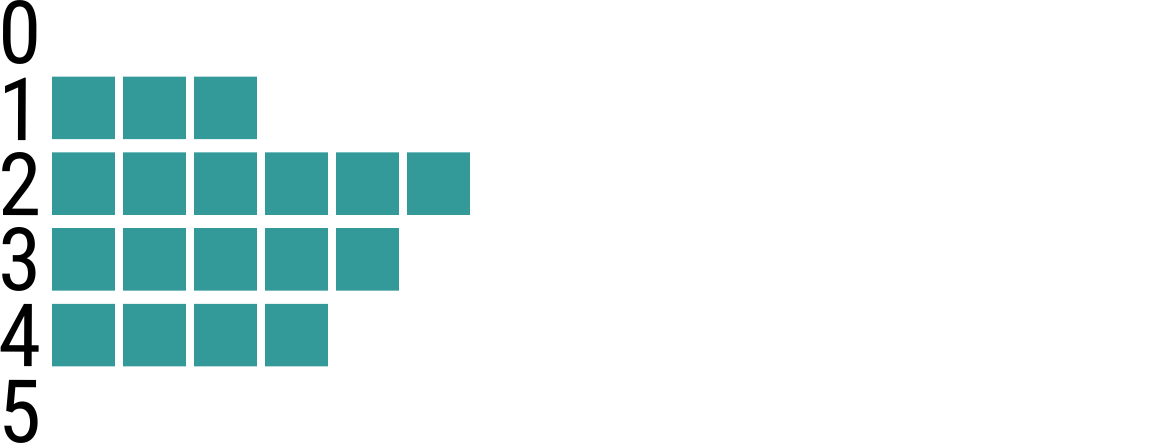} &
        \includegraphics[width = 2.25cm, valign=c]{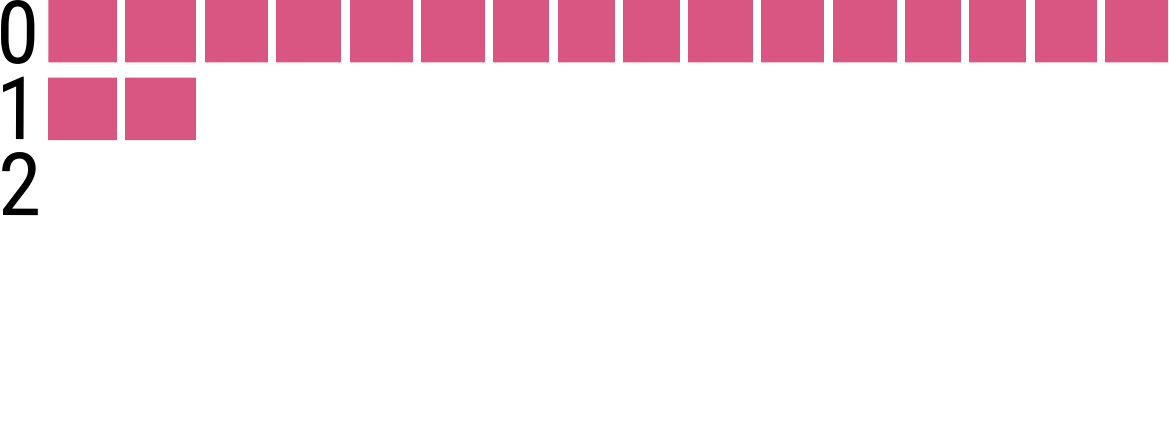} &
        \includegraphics[width = 0.9cm, valign=c]{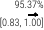} &
        \includegraphics[width = 0.9cm, valign=c]{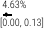} \\ 
        \midrule

        \includegraphics[height = 0.6cm, valign=c]{Figures/icons/donut.pdf} &
        486 & 
        \includegraphics[width = 2.25cm, valign=c]{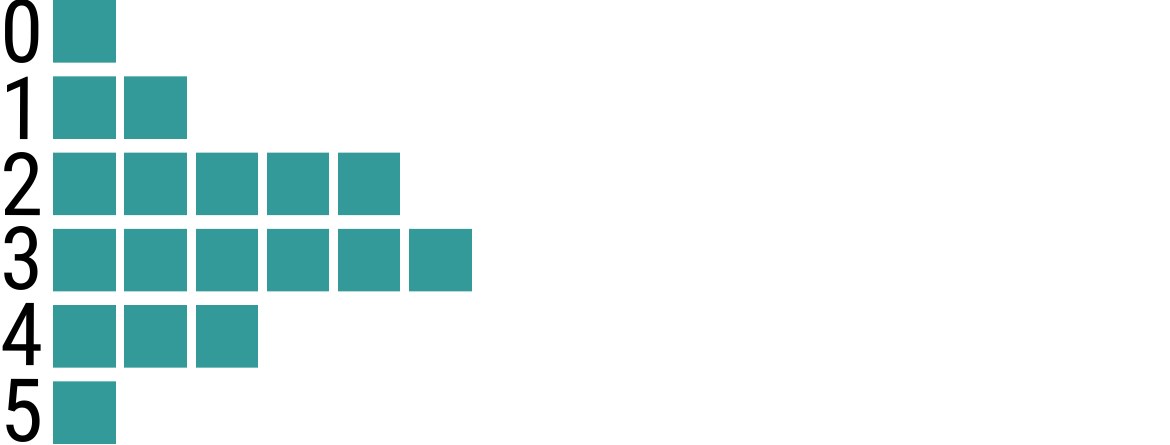} &
        \includegraphics[width = 2.25cm, valign=c]{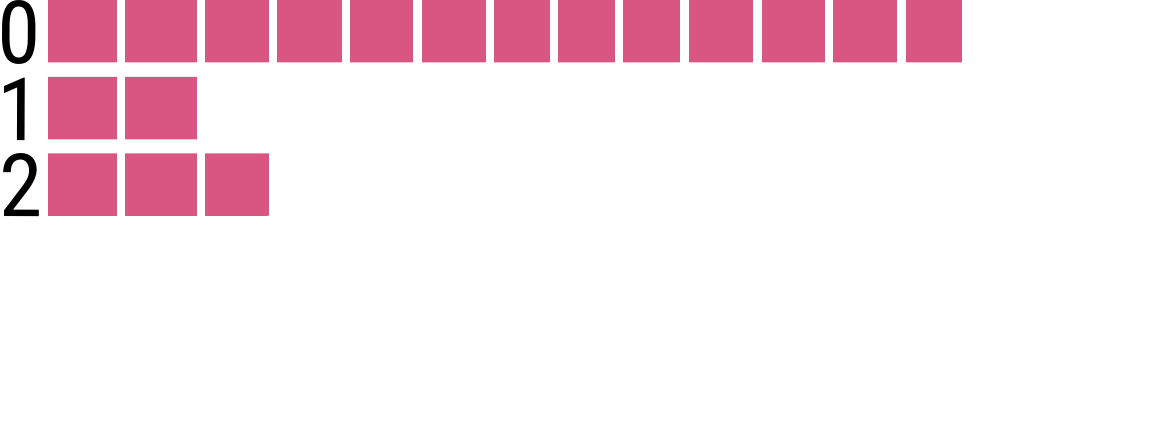} &
        \includegraphics[width = 0.9cm, valign=c]{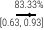} &
        \includegraphics[width = 0.9cm, valign=c]{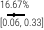} \\ 
        \midrule

        \includegraphics[height = 0.6cm, valign=c]{Figures/icons/texture.pdf} &
        432 & 
        \includegraphics[width = 2.25cm, valign=c]{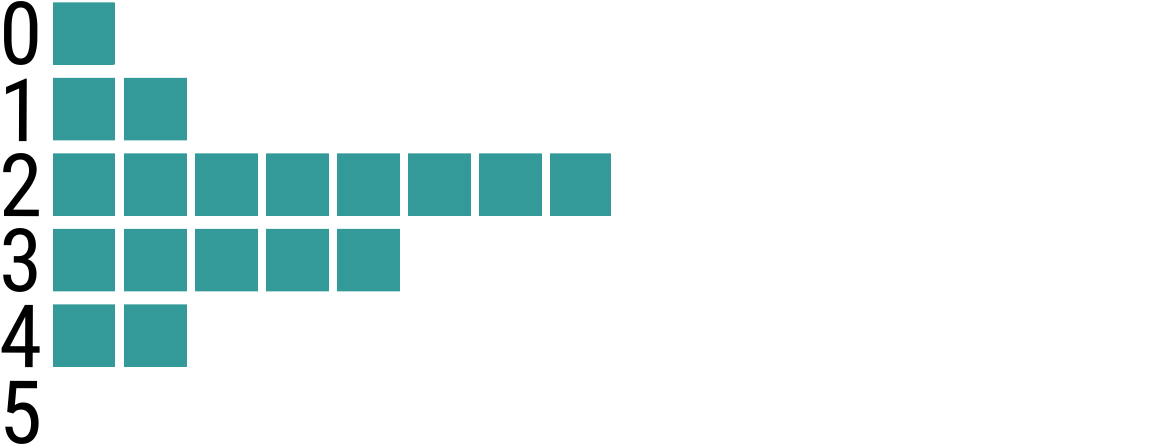} &
        \includegraphics[width = 2.25cm, valign=c]{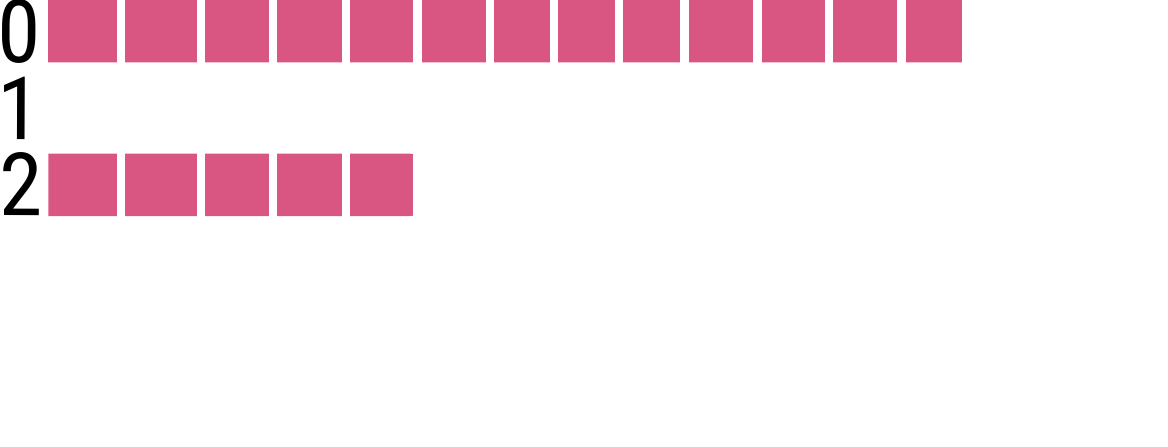} &
        \includegraphics[width = 0.9cm, valign=c]{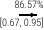} &
        \includegraphics[width = 0.9cm, valign=c]{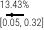} \\ 
        \bottomrule
        
    \end{tabu}
    \label{tab:gamingperformance}
    \vspace{-5mm}
\end{table}

\section{An Evaluation of Visualizations in Motion:\\ Study Results}
Our participants were engaged in playing \emph{RobotLife} as they made a clear effort to win the game (\autoref{tab:gamingperformance}).
The majority of our participants did not lose a single time with all visualization designs.
A detailed statistical comparison can be found at Appendix-\autoref{tab:performanceStatisticalAnalysis}. While there is no strong evidence of a difference between designs, there may be a trend for the \texttt{non-integrated design} to have a higher win ratio. On average, participants lost the least times and had the highest win ratio (wins over all played games) with the \includegraphics[height = 0.32cm, valign=c]{Figures/icons/bar.pdf}\ \texttt{non-integrated design}. Participants killed the most evil robots, played the most games (won and lost), but also had the highest loss ratio (losses over all games played) with the \includegraphics[height = 0.32cm, valign=c]{Figures/icons/donut.pdf}\ \texttt{partial-match design}.  
Participants killed the least robots with the \includegraphics[height = 0.32cm, valign=c]{Figures/icons/texture.pdf}\ \texttt{fully-integrated design}.
Next, we report the features that players reported to have affected their gaming performance, followed by the user experience of each visualization in motion design. 

\subsection{Features that Affected Gaming Performance}
To clarify what relevant factors affected participants' gaming experience, we qualitatively analyzed their answers to open-ended post-interview questions through open coding. Our coding was initially done independently by two authors and then merged together. The coding details can be found in supp.\ material.

\vspace{1pt}
\noindent\textbf{Motion Factors:}
Participants explained how their character's movement, which was under their control, and other robots, which moved autonomously and were out of control, affected their performance.

\noindent\textit{Under control movement:}
\percentagebar{0.22}4/18 participants stated that they needed to stop or slow down their gameplay to accurately read proportions, particularly when values approached 66\% (the threshold for identifying evil robots). Notably, two of them highlighted that this interruption significantly hampered their performance, impeding quick action.

\noindent\textit{Autonomous Movement:}
\percentagebar{0.11}4/18 participants found that the movement out of their control had a negative impact. Specifically, the autonomous motion could lead to a visualization deformation (\texttimes\ 2p) and the change of viewing angles (\texttimes\ 2p), such as rotation of a character and its attached visualization, which made reading challenging, particularly when the visualization was not a traditional design like a bar.  
Seven participants reported being used to moving characters in FPS games and, as such, could be unaffected by their autonomous movement.

\vspace{1pt}
\noindent\textbf{Contextual factors:}
Over half of our participants \percentagebar{0.61}11/18 stated that the contextual factors, such as the busy background and the lighting, did not distract them. They explained that they were used to such factors and naturally ignored these factors to focus only on the primary tasks. In contrast, the remaining participants, who answered, commented on the impact of color and lighting conditions on their performance.

\noindent\textit{Color:}
\percentagebar{0.22}4/18 participants appreciated the high color contrast between the visualization (red charts) and the game world (dark green), which aided them in quicker robot identification and task completion. No participant mentioned the difference in color intensity for \textureS\ caused by the overlay on the robot skin.

\noindent\textit{Lighting:}
\percentagebar{0.06}1/18 participant reported that the lighting distracted them due to the slight color changes on robots' skin rendering from different views, especially for \textureS. 

\vspace{1pt}
\noindent\textbf{Reading strategies:}
Almost all participants \percentagebar{0.94}17/18 reported that they found a \emph{visual referent}, which could help them to quickly and precisely identify the threshold. They defined the visual referent as a specific fixed position on the game character that could indicate a percentage close to 66\%, such as the robots' eye position, a particular locations on the eye, or specific lines on the robots' texture. 
This finding echoes prior work for reading bar and donut charts in motion where participants reported ``dividing a chart into several parts'' \cite{Yao:2022:VisinMotion}. Two participants also reported comparisons between robots: When a robot had a tricky health value that was difficult to determine, they would leave the robot and move around to find another one to compare.

\vspace{1pt}
\noindent\textbf{In summary}, several features apart from the visualization type affected players' perceived performance. 
When the gaming tasks related to reading from visualizations in motion, high contrast colors and proper visual referents helped players, while motion factors might impede visualization reading and slow down players.

\begin{table*}[tb]
    \centering
    \captionsetup{justification=centering}
    \footnotesize
    \caption{Right: Measured aesthetics and readabilities from the in-game questionnaire on 7-point Likert scales.\\Left: Rankings of visualization designs according to the questions asked in the post-questionnaire.\\%
    \includegraphics[height = 0.32cm, valign=c]{Figures/icons/bar.pdf}: \texttt{non-integrated design}, %
    \includegraphics[height = 0.32cm, valign=c]{Figures/icons/donut.pdf}: \texttt{partial-\-match design}, %
    \includegraphics[height = 0.32cm, valign=c]{Figures/icons/texture.pdf}: \texttt{fully-integrated design}
    }%
    \tabulinesep=0mm
    \aboverulesep = 0.5mm
    \belowrulesep = 0.5mm
    \begin{tabu}{@{\hspace{0ex}}C{1cm}@{\hspace{0.5ex}}
                    |@{\hspace{0.5ex}}C{1.8cm}@{\hspace{0.5ex}}
                    |@{\hspace{0.5ex}}C{1.8cm}@{\hspace{0.5ex}}
                    |@{\hspace{0.5ex}}C{1.8cm}@{\hspace{0.5ex}}
                    |@{\hspace{0.5ex}}C{1.8cm}@{\hspace{0.5ex}}
                    ||@{\hspace{0ex}}C{1.75cm}@{\hspace{0ex}}
                    |@{\hspace{0ex}}C{1.75cm}@{\hspace{0.5ex}}
                    |@{\hspace{0ex}}C{1.75cm}@{\hspace{0.5ex}}
                    |@{\hspace{0ex}}C{1.75cm}@{\hspace{0.5ex}}
                    |@{\hspace{0.5ex}}C{1.8cm}@{\hspace{0ex}}
                    }
        \toprule
        \includegraphics[width = 1cm, valign=c]{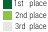} &
        Readability under motion & 
        Aesthetic Fit &
        Task Support &
        Overall Experience & 
        \emph{Aesthetic} & 
        \emph{Clear} &
        \emph{Readable} & 
        \emph{Intuitive} & 
        \emph{Understandable}
        \\
        \midrule
        
        \includegraphics[height = 0.8cm, valign=c]{Figures/icons/bar.pdf} &
        \includegraphics[width = 1.8cm, valign=c]{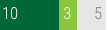} &
        \includegraphics[width = 1.8cm, valign=c]{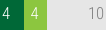} &
        \includegraphics[width = 1.8cm, valign=c]{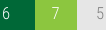} & 
        \includegraphics[width = 1.8cm, valign=c]{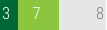} & 
        \includegraphics[width = 1.75cm, valign=c]{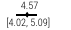} & 
        \includegraphics[width = 1.75cm, valign=c]{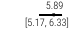} &
        \includegraphics[width = 1.75cm, valign=c]{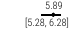} &
        \includegraphics[width = 1.75cm, valign=c]{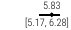} &
        \includegraphics[width = 1.75cm, valign=c]{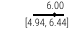}\\ 
        \midrule

        \includegraphics[height = 0.8cm, valign=c]{Figures/icons/donut.pdf} &
        \includegraphics[width = 1.8cm, valign=c]{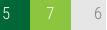} &
        \includegraphics[width = 1.8cm, valign=c]{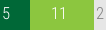} &
        \includegraphics[width = 1.8cm, valign=c]{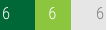} & 
        \includegraphics[width = 1.8cm, valign=c]{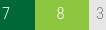} & 
        \includegraphics[width = 1.75cm, valign=c]{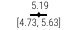} &
        \includegraphics[width = 1.75cm, valign=c]{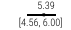} &
        \includegraphics[width = 1.75cm, valign=c]{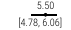} &
        \includegraphics[width = 1.75cm, valign=c]{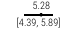} &
        \includegraphics[width = 1.75cm, valign=c]{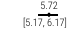} \\ 
        \midrule

        \includegraphics[height = 0.8cm, valign=c]{Figures/icons/texture.pdf} &
        \includegraphics[width = 1.8cm, valign=c]{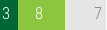} &
        \includegraphics[width = 1.8cm, valign=c]{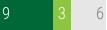} &
        \includegraphics[width = 1.8cm, valign=c]{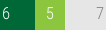} & 
        \includegraphics[width = 1.8cm, valign=c]{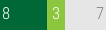} & 
        \includegraphics[width = 1.75cm, valign=c]{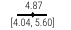} & 
        \includegraphics[width = 1.75cm, valign=c]{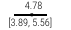} &
        \includegraphics[width = 1.75cm, valign=c]{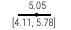} &
        \includegraphics[width = 1.75cm, valign=c]{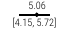} & 
        \includegraphics[width = 1.75cm, valign=c]{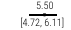}\\ 
        \bottomrule
        
    \end{tabu}
    \label{tab:postquestionnaire}
    \vspace{-6mm}
\end{table*}

\subsection{User Experience of Visualization Designs}
To understand the user experience of our visualization designs, we analyzed two aspects: in-game Likert scales on aesthetics and readability (\emph{clear}, \emph{readable}, \emph{intuitive}, and \emph{understandable}) of each design and a qualitative analysis of participants' design preferences.

We conducted quantitative analysis of Likert scales. The BeauVis scale \cite{He:2023:BeauVis} that we used is a validated scale for aesthetic pleasure measurement of visualization. We thus report a single aesthetic score per visualization design. In contrast, we individually report results for each readability Likert item as it has not been validated yet as a complete scale. We present means and 95\% confidence intervals (CIs), indicating our confidence that the population mean falls within this interval. Additionally, we compare different conditions using CIs of mean differences, a method utilized in Likert scale analysis in previous visualization research\cite{Dragicevic:2018:Blinded,Dragicevic:2019:Multiverse}. We generated all CIs using Bias-Corrected and accelerated (BCa) bootstrapping (10,000 iterations) and analyzed them using estimation techniques, as recommended in the literature \cite{cockburn2020threats,cumming2013understanding,dragicevic2016fair}. A non-overlap of CIs with 0 in mean differences suggests evidence of a difference, akin to statistically significant results in traditional \textit{p}-value tests. \textit{p}-values equivalent to CI results can be derived following Krzywinski and Altman~\cite{krzywinski2013points}. Means and CIs are shown in \autoref{tab:postquestionnaire}-Right. Pairwise differences and CIs are in Appendix-\autoref{tab:statisticalAnalysis}.

Our qualitative analysis was again conducted through coding by two authors individually, based on the ranking participants made and the reasons they explained to each question asked in the post-questionnaire. Results are shown in \autoref{tab:postquestionnaire}-Left. 
Since our paper focuses on the design of visualizations in motion, we categorize and report below our analysis and results based on each visualization design.

\subsubsection{\barIcon\ Non-integrated design}
\noindent\textbf{Measured aesthetics \& Aesthetic fitting:} 
\BarS\ was measured as the least aesthetic design. Participants ranked it last regarding how it fit our game character and world. \BarS\ was referred to as \emph{less immersive} and \emph{somehow artificial} due to the visual separation to its data referent.

\vspace{1pt}
\noindent\textbf{Measured readabilities \& Readability under motion:} Regarding readability, \barS\ was measured as the most clear, readable, intuitive, and understandable design. Statistically, we see evidence that \barS\ was clearer and more intuitive than \textureS. \BarS\ was ranked as the most readable design under motion with votes over 3\texttimes\ higher than \textureS\ and 2\texttimes\ higher than \donutS. Participants explained that \barS\ is a \emph{typical visual representation} with a \emph{common embedding location} in FPSs. They were thus familiar with such designs and well-trained to read them even under motion. Participants commented that the embedding location of \barS\, that was over the head of the character, was used as a \emph{visual indicator} to identify where the robot was, even in a moving and crowded environment.   

\vspace{1pt}
\noindent\textbf{Task support:}
All three designs received the same votes for the 1\ts{st} place of game task support. \BarS\ got the most votes for the 2\ts{nd} place and the least votes for the last place. The \percentagebar{0.72}13/18 participants who voted for the 1\ts{st} and 2\ts{nd} places explained that the game tasks were highly relevant to their reading results. Thus, they answered the question of task support based on the designs' readability under motion. The remaining 5 participants shifted their ranking orders to find a balance between easy-to-read and aesthetic-fit.

\vspace{1pt}
\noindent\textbf{Overall experience:}
\BarS\ was ranked lowest in terms of overall experience. Most participants commented that \barS\ was a \emph{boring} visualization since it is too typical in video games and \emph{less innovative}.

\vspace{1pt}
\noindent\textbf{In summary,}
\barS\ is highly readable for moving robots because it was visible even behind occluding objects. Its embedding location (over the data referent) helped to find robots, but participants found it less aesthetically pleasing and thought it was less immersive. Its typical visual representation requires less training, experience, and effort to read but also lacks innovation and might disappoint players who look forward to a novel user experience.

\subsubsection{\donutIcon\ Partial-match design}
\noindent\textbf{Measured aesthetics \& Aesthetic fitting:} 
\DonutS\ was measured as the most aesthetic design and statistically more aesthetic than \barS. It was ranked at the 2\ts{nd} place in terms of best fitting our game context. Participants said that \donutS\ was not a widely used design in FPSs and was \emph{somehow innovative}. They reported that \donutS's representation matched part of its referent's shape and was thus \emph{less artificial}. Participants commented that although the embedding location of \donutS\ overlapped with its referent and \emph{seemed immersive}, it still floated above the referent from some perspectives.

\vspace{1pt}
\noindent\textbf{Measured readabilities \& Readability under motion:} 
\DonutS\ was measured in the middle for questions on clear, readable, intuitive, and understandable and was ranked in the 2\ts{nd} place for reading under motion. Participants did not give specific comments regarding the readability of \donutS.

\vspace{1pt}
\noindent\textbf{Task support:} 
\DonutS\ received fairly equal votes for the 1\ts{st}, the 2\ts{nd}, and the 3\ts{rd} place of game task support. Participants stated again that their task completion depended on reading results. Since they had a clear preference for the most and the least readable design, \donutS\ was put in between as a \emph{trade-off option} by considering its aesthetic fit to the game context.

\vspace{1pt}
\noindent\textbf{Overall experience:}
The majority of participants (\percentagebar{0.83}15/18) put \donutS\ at the 1\ts{st} or the 2\ts{nd} place of a design that gives positive overall experience. 7 participants appreciated \donutS\ because its visual representation could fit well with the partial shape of its referent. Another 3 participants said that they were not that used to reading a circular bar. The visual representation was actually an \emph{extra challenge}. Interestingly, 2 of the 3 participants expressed that they preferred challenging games more than easy-to-win games, for a greater sense of accomplishment.

\vspace{1pt}
\noindent\textbf{In summary,}
participants did not have a clear preference for \donutS\ and always treated it as a trade-off option to balance overall considerations. The embedding location of \donutS\ (overlap with data referent) could give a somewhat immersive experience. The visual representation of \donutS\ may excite players at some level, bringing both task challenge and game attractiveness.

\subsubsection{\textureIcon\ Fully-integrated design}
\noindent\textbf{Measured aesthetics \& Aesthetic fit:} 
\TextureS\ was measured in the middle for aesthetics but ranked in 1\ts{st} place for best fit our game context aesthetic, with approximately 2\texttimes\ more votes than other designs. Participants who liked \textureS\ reported that this design was \emph{realistic} and \emph{immersive} because it was fully coupled with its data referent. In contrast, participants who disliked it complained that the full integration \emph{broke the design} of the game character, especially when the percentage was extremely high --- the entire robot looked red, which annoyed them. These participants wanted to see the characters' skin and attached gaming features. They explained that game characters are always well-decorated in commercial games and deserve to be enjoyed.

\vspace{1pt}
\noindent\textbf{Measured readabilities \& Readability under motion:}
According to the rating scales, \textureS\ was the least clear, readable, intuitive, and understandable design. It was also ranked as the least readable design under motion. Participants stated that they had to be very careful when reading from \textureS\ due to the dynamic design of robots: sometimes the robot became shorter and sometimes taller. The inconsistent height of the robot and its shape deformation added \emph{additional difficulties} for participants' estimation, as the visualization was embedded in the entire robot.

\vspace{1pt}
\noindent\textbf{Task support:} 
\TextureS\ received the most votes as the least task-supporting design. Participants commented that although \textureS\ was quite a new design in FPSs and could integrate well in our game context, its poor readability could not support the reading task, especially when heavy motion involved due to many robot movements and dynamic shape changes.

\vspace{1pt}
\noindent\textbf{Overall experience:}
\textureS\ received the most votes (\texttimes\ 8p) for the 1\ts{st} place for a positive overall experience, though it was the least readable and task-supporting design. These 8 participants thought that \textureS\ \emph{ well} in the game world. Three of them stated they had never seen such a design in practice and were excited to see \emph{new} in-game visualizations. Participants also commented that their performance with \textureS\ was worse than their usual performance. They thought it might be due to unfamiliarity with such a visualization and its embedding method, but it could be overcome after enough training. Participants wanted to see more innovative visual representation designs, especially in games, even if it might initially hurt their scores.

\vspace{1pt}
\noindent\textbf{In summary,}
\textureS\ has a good integration in our game context, giving players an immersive experience. However, it may not be ideal for reading-based tasks, especially with dynamic changes and multiple types of motion. The embedding location of \textureS\ ( into the referent) is rare in FPSs which enhances realism but also brings extra challenges for players due to its high integration level. The visual representation of \textureS\ could not be seen everywhere. Its rarity, however, had a novelty effect and brought a unique gaming experience.

\subsubsection{Summary}
In our evaluation, each design had specific advantages and shortcomings. Not only do the embedding location, visual representation, and visual encoding affect user experience, but the motion conditions, the context, and the tasks the visualization supported also play a significant role. Rather than simply saying which design was best, there are trade-offs to consider based on concrete scenarios and user needs.

\subsection{General comments}
Ten participants left general comments. Nine of them commented that (a) our game was really cool, extremely fun, and super well made; (b) our study was interesting and a bit challenging, specifically for \textureS\ condition.
One participant, who works in the game industry, mentioned being encouraged to start considering trade-offs when practically designing in-game visualizations in motion.

\section{Considerations for Design and Future Research}
Our work is a first exploration of how users experience visualizations in motion in a concrete application context: video games. An important outcome of exploratory research are open research avenues and design considerations. We discuss these here by drawing on our systematic review and evaluation results.
We first discuss how our results relate to the game context and then how they may relate to other contexts.

\subsection{In-game visualization in motion design}
\noindent\textbf{Standards, diversity, and simplicity:} 
Our systematic review revealed many instances of visualizations in motion in video games. Quantitative and qualitative data was mainly represented by bar charts and text labels, and categorical data by text labels or signs. Visual representations were simple or replaced by text and often showed a single data dimension. Past work on the glanceability of micro visualizations \cite{Blascheck:2023:ParttoWhole}  (not under motion) showed that text was harder to read quickly compared to bar or donut charts. Similarly in games, players can only afford quick glances to make decisions during their gameplay. This begs the question: Are bars and text an expected standard in games, or are game designers reluctant to add different visualizations because they are unsure how players perceive and perform with them? Is text perhaps less effective than expected in video games as well? Further research on which representation is effective for which game data would be useful.

\vspace{1pt}
\noindent\textbf{Visibility and blending in the game:} 
Several gaming situations may compromise the readability of visualizations due to movement. For example, overlap between moving game entities, like multiple characters and their attached visualizations, is common. Moreover, a very subtle visualization such as \textureS\ could improve immersion but may also blend too much into the background and no longer be as visible as explicit representations that visually pop out. It remains an open question to determine to what extent players are willing to compromise the readability/visibility of visualizations in motion and how to improve the perception of in-game visualizations in motion in such visibility-limited situations. 

\noindent\textbf{Impact of visualization familiarity on game-play:} Several well-trained players commented that they performed better on \barS\ because they were used to this typical design. They thought their game performance with \textureS\ would be improved if they were more familiar with it.
The effectiveness of learning to read in-game visualizations in motion certainly requires further study. In addition, knowing the limits when even well-trained viewers can no longer read in-game visualizations in motion could aid video game visualization designers.

\subsection{Visualizations in motion in concrete contexts}
\noindent\textbf{Aesthetics:}
Aesthetics are important for visualizations in motion embedded in a context to visually connect to the referent and context. Two important choices need to be made: (a) To what extent should the visualization's color stand out from the remainder of the context? (b) How closely should a visual representation follow its referent's shape? 
For an immersive and realistic experience, the visualization should be embedded closely to its referent. Yet, our study showed clear tradeoffs between aesthetic preferences and perceived effectiveness. The readability of a visual representation might be assessed on a design-by-design basis.

\noindent\textbf{Glanceability:}
There might be a tradeoff between canonical visualizations and embedding locations. Well-known embedding locations could help viewers quickly locate the visualization and give them some extra time to read. Common representations may also speed up reading under the tradeoffs mentioned above. To support glanceability, visualizations in motion should practically represent simple and single pieces of information rather than complex data. However, further studies are required to determine the extent of possible complexity.

\noindent\textbf{Expectations:}
Our game task required the correct perception of visualizations in motion, but participants did not necessarily favor the highly readable design. Instead, several participants would prefer to degrade their regular game performance in favor of novel visualization designs, even with lower task support. This finding is a reminder that viewers may sometimes have hidden expectations, such as innovation. What these expectations might be in diverse scenarios needs further research.

\subsection{Visualization in motion in other contexts}
Our results are derived from studying visualizations in an FPS game. However, there are other game genres in which game characters or avatars play a central role (\eg, role-playing games and exercise games). For these, we expect many of our considerations also to hold, ensuring correct and quick readability but also supporting aesthetic appeal or familiarity, for example, remains crucial. This goal might be more important in games with a very large number of game characters, like team sports games (\eg, FIFA).
Visualizations in motion may also be applied outside of the video game context, like in sports videos \cite{Lin:2020:SportsXR,Lin:2021:ARVisforBasketballTraining,Lin:2022:Omnioculars,Yao:2022:VisForSwimming}. For swimming videos,  Yao et al.\ \cite{Yao:2022:VisForSwimming} found a prevalence of small and simple representations are often be shown in text. Islam et al. observed \cite{Islam:2020:VisOnWatchFaces, Yao:2022:FitnessTrackers} that visualizations may be used on smartwatch faces instead of text and may be read in motion while walking or running. Yet, it is still unclear how simple visualizations need to be for a motion context or whether text is actually an effective type of representation here. In fact, a study on smartwatches (albeit without motion) found that text was less effective than donut or bar charts \cite{Blascheck:2023:ParttoWhole}. An open question is how the requirement for visualizations to ``blend in'' vs.\ ``stand out'' may be context-dependent. 
Will ``blending in'' be more important when viewers expect to experience a game world than read visualizations in a TV broadcast? 
Finally, video games have one main advantage over embedded visualizations in motion in real-time sports scenarios: they typically allow players to be trained to learn to read a visualization. Onboarding viewers of a live broadcast will have very different challenges to consider.
As such, we call for more research on visualization in motion in concrete contexts. Studying multiple contexts will allow us to learn and compare considerations and challenges and ultimately derive overarching and concrete recommendations.

\section{Conclusion}
Our research explores the user experience and design guidelines of visualization in motion in a concrete context. We systematically reviewed how embedded visualizations in motion have been designed in video games. Our review highlighted several visualization designs. 
To explore some possible tradeoffs in the user experience of in-game visualizations in motion, we conducted a user study based on our FPS game, \emph{RobotLife}.
Our results showed diverse design preferences and clear trade-offs between visualization readability, perceived aesthetics, task support, and overall experience. We also proposed a set of dedicated design considerations as well as experimentation opportunities for visualization in motion. 

\acknowledgments{%
This work was partly supported by the Agence Nationale de la Recherche (ANR), grant number ANR-19-CE33-0012. We thank Micka{\"e}l Sereno for help with implementing the robot texture.
}

\section*{Images/graphs/plots/tables/data license/copyright}
All figures, icons, and tables in this paper are under our own copyright or released under the Creative Commons CC BY-SA license \ccLogo \ccAttribution \ccShareAlike, except \autoref{fig:embeddinglocations}. 

\section*{Supplemental Material Pointers}
The pre-registrations for our user study can be found at (a) \href{https://osf.io/h4cks}{\texttt{https://osf.io/h4cks}}.
Respectively. We also share our (a) systematic review material, results, and scripts; (b) scripts for the analysis of the Likert scales and participants' performance; (c) user study demonstration video; and (d) an executable RobotLife application and source codes of \emph{RobotLife} and other materials (appendix, ethics approval...) at \href{https://osf.io/3v8wm/}{\texttt{osf.io/3v8wm/}}.

\bibliographystyle{abbrv-doi-hyperref}
\bibliography{visinmotion}

\appendix 
\clearpage

\setcounter{section}{3}
\setcounter{subsection}{1}
\subsection{Categorization of Current In-game Visualizations}
Among our 160 collected in-game visualizations in motion, the genres with the most occurrences were Role-playing Games (RPG), with 18/160 visualizations; strategy games, with 16/160; and wargames, with 14/160 samples. We categorized these visualizations based on multiple dimensions related to situated visualization and motion characteristics. The coding of categorization was done by two authors, A and B, and reviewed by another author, C. Following, we describe our coding process in detail:
\begin{enumerate}[label=(\alph*)]
    \item  During the first round of coding, A and B discussed broad categories together and coded individually all collections according to these categories. Meanwhile, A and B marked “difficult” collections separately for those collections that they thought could be categorized into several categories and required a new category. A and B finished categorizing all collections and moved to the second round of coding.
    \item In the second round of coding, A and B verified for each collection if they had put it in the same category. If not, A and B discussed the collection to form a joint opinion and put the discussed collection into an existing or a new category. Meanwhile, they added annotations to all discussed collections and noted the reasons for their decision.  A and B repeated these steps until they finished with all collections. 
    \item In the third round of coding, C reviewed the categories provided by A and B, including all sub-collections and annotations. C asked and discussed with A and B when C did not agree with the category of a collection. A, B, and C would start a new discussion to determine where a controversial collection should be categorized. A, B, and C repeated these steps until all coders were satisfied with the final coding results.
\end{enumerate}

\begin{table*}[h]
\centering
\caption{50 video games selected from 17 genres}
\scriptsize
\begin{tabular}{| l | l | l | l | l |}
\rowcolor{bluePoli}
\hline
 &  \color{white}\textbf{Genre} &  \color{white}\textbf{Video game Title} &  \color{white}\textbf{Platform} &  \color{white}\textbf{Year} \\    \hline \hline
1 & Action & Gear of war & Xbox One & 2016 \\
2 & Action & Call of Duty: Black Ops 4 & PlayStation 4 & 2018 \\
3 & Action & Battlefield V & PlayStation 4 & 2018 \\
4 & Adventure & Detroit: Become Human & PlayStation 4 & 2018 \\
5 & Adventure & Subnautica: Below Zero & Xbox Series S & 2021 \\
6 & Adventure & Sam and Max Save the World - Remastered & PC & 2020 \\
7 & Fighting & Super Smash Bros & Wii U & 2014 \\
8 & Fighting & Deadpool &  PC & 2013 \\
9 & Fighting & PlayStation All-Stars Battle Royale & PlayStation 3 & 2012 \\
10 & FPS & RAGE 2 & PlayStation 4 & 2019 \\
11 & FPS & Borderlands 3 & PlayStation 4 & 2019 \\
12 & FPS & Halo Infinite & Xbox Series X & 2021 \\
13 & Flight/Flying & Microsoft Flight Simulator & Xbox Series S & 2021 \\
14 & Flight/Flying & The Falconeer & PC & 2020 \\
15 & Flight/Flying & Skies of Fury DX & Switch & 2018 \\
16 & Party & Mario Party 9 & Wii & 2012 \\
17 & Party & Nintendo Land & Wii U & 2012 \\
18 & Party & Go Vacation & Wii & 2011 \\
19 & Platformer & Donkey Kong Country: Tropical Freeze & Wii U & 2014 \\
20 & Platformer & Super Mario 3D World & Wii U & 2013 \\
21 & Platformer & Terraria & PC & 2011 \\
22 & Racing & Mario Kart Tour & iOS & 2019 \\
23 & Racing & Gran Turismo 7 & PlayStation 5 & 2022 \\
24 & Racing & Forza Horizon 5 & Xbox Series S & 2021 \\
25 & RTS & Orcs Must Die! 3 & PC & 2021 \\
26 & RTS & Pikmin 3 Deluxe & Switch & 2020 \\
27 & RTS & Crusader Kings III & PC & 2020 \\
28 & RPG & Cyberpunk 2077 & PC & 2020 \\
29 & RPG & Wolcen: Lords of Mayhem & PC & 2020 \\
30 & RPG & Biomutant & PlayStation 4 & 2021 \\
31 & Simulation & MechWarrior 5: Mercenaries & PC & 2019 \\
32 & Simulation & Dirt 5 & PlayStation 5 & 2020 \\
33 & Simulation & F1 2020 & PlayStation 4 & 2020 \\
34 & Sport & NBA 2K21 & PlayStation 5 & 2020 \\
35 & Sport & Madden NFL 21 & PlayStation 4 & 2020 \\
36 & Sport & MBL The Show 20 & PlayStation 4 & 2020 \\
37 & Strategy & Phoenix Point & PC & 2019 \\
38 & Strategy & Gears Tactics & PC & 2020 \\
39 & Strategy & A Total War Saga: TROY & PC & 2020 \\
40 & TPS & Aliens: Fireteam Elite & PC & 2021 \\
41 & TPS & Returnal & PlayStation 5 & 2021 \\
42 & TPS & Outriders & PC & 2021 \\
43 & Turn-Based Strategy & King’s Bounty II & PlayStation 5 & 2021 \\
44 & Turn-Based Strategy & Sakura Wars & PlayStation 4 & 2020 \\
45 & Turn-Based Strategy & XCOM: Chimera Squad & PC & 2020 \\
46 & Wargames & Worms Battlegrounds & PlayStation 4 & 2014 \\
47 & Wargames & Toy Soldiers & PC & 2012 \\
48 & Wargames & Sengoku & PC & 2011 \\
49 & Wrestling & WWE 2K14 & PlayStation 3 & 2014 \\
50 & Wrestling & Fire Pro Wrestling & Xbox 360 & 2012 \\
\hline
\end{tabular}
\label{table:selectedGames}
\end{table*}
\clearpage

\label{sec:Categorization}
\begin{figure*}[tb]
    \centering
    \includegraphics[width=\linewidth]{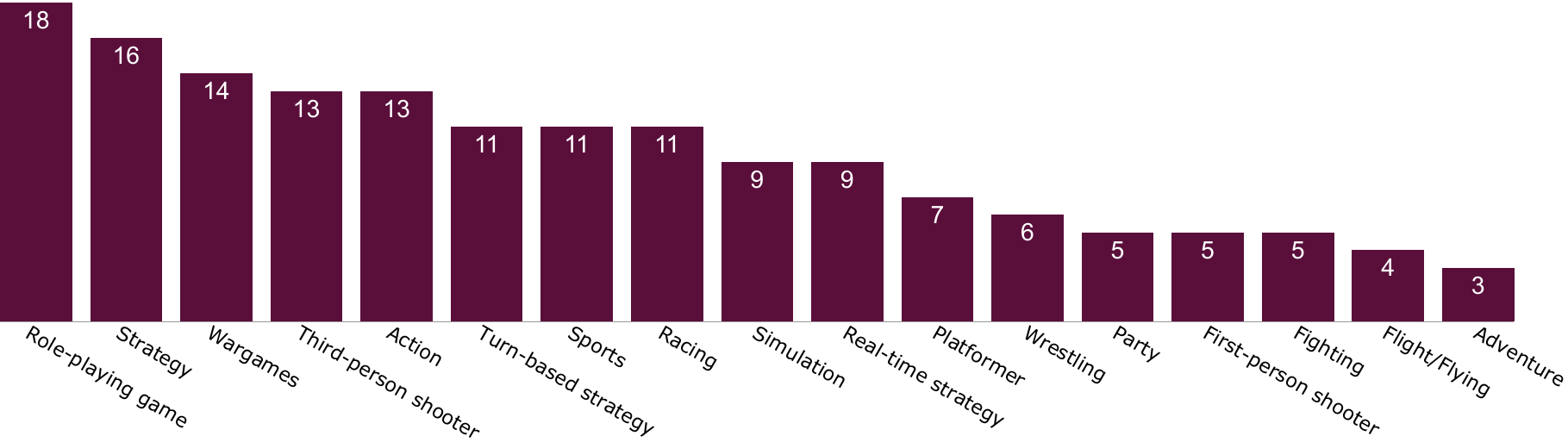} 
    \caption{Counts of 160 collected visualizations in motion per game genre}
    \label{fig:genres}
\end{figure*}
\clearpage

\begin{figure*}[t]
    \centering
    \includegraphics[width=\linewidth]{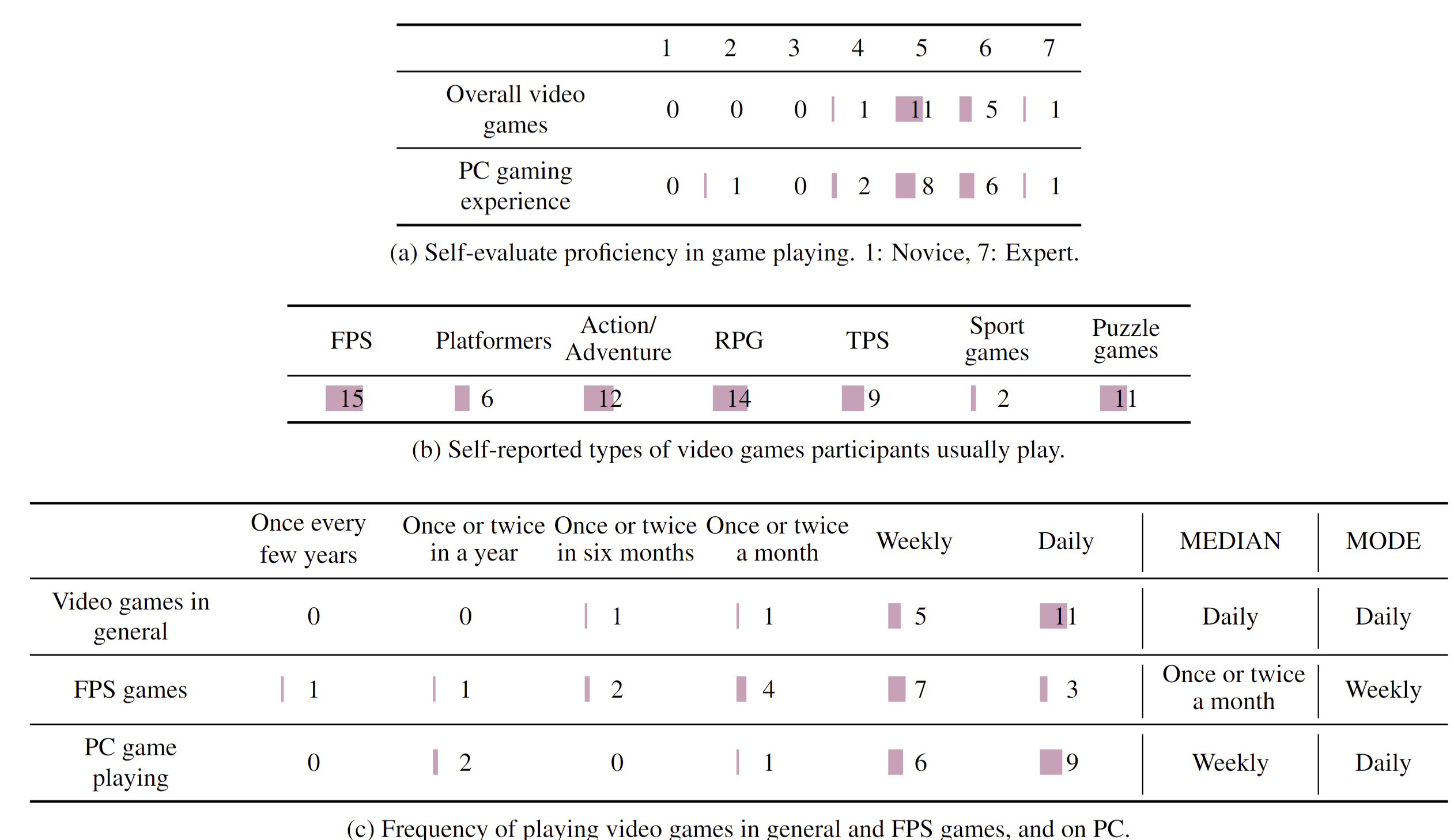} 
    \caption{Gaming experience, games familiarity, and experience in playing games on PC of our participants in formal study.}
    \label{tab:gamefamilarity}
\end{figure*}
\clearpage

\setlength{\picturewidth}{.4\linewidth}
\setlength{\pictureheight}{2.2cm}
\begin{table*}
\renewcommand{\arraystretch}{1.7}  
    \caption{Detailed results of measured aesthetics and readabilities. Top: Mean and pairwise differences between conditions of the 7-point Likert scale on aesthetics. Bottom: Means and pairwise differences between conditions of four Likert items on the 7-point Likert scale on readability.
    Error bars represent 95\% Bootstrap confidence intervals (CIs) in black, adjusted for pairwise comparisons with Bonferroni correction (in red). In the Pairwise difference image, CIs that do not cross 0 indicate evidence of a difference between the two techniques -  the tighter the CI of the difference and the furthest from 0, the higher the evidence of a difference. 
    Our results provide clear evidence that the \includegraphics[height = 0.32cm, valign=c]{Figures/icons/donut.pdf}  \texttt{partial-\-match design} was perceived as having higher aesthetics than the \includegraphics[height = 0.32cm, valign=c]{Figures/icons/bar.pdf} \texttt{non-integrated design} but no evidence of difference between the others in terms of aesthetics. There is evidence that the \includegraphics[height = 0.32cm, valign=c]{Figures/icons/texture.pdf}  \texttt{fully-integrated design} was less clear and less intuitive than the  \includegraphics[height = 0.32cm, valign=c]{Figures/icons/bar.pdf} \texttt{non-integrated design} and possibly a trend that it is also less readable.  }
    \label{tab:statisticalAnalysis}
    \scriptsize
    \begin{tabular}{@{}p{1.97\picturewidth}@{}
                    @{}r@{~}
                    @{}p{1.8cm}@{}
                    @{}p{1.87\picturewidth}@{}
                    @{}r@{~}
                    @{}p{1.5cm}@{}
                    }
    \multirow{3}{*}{\includegraphics[height=\pictureheight]{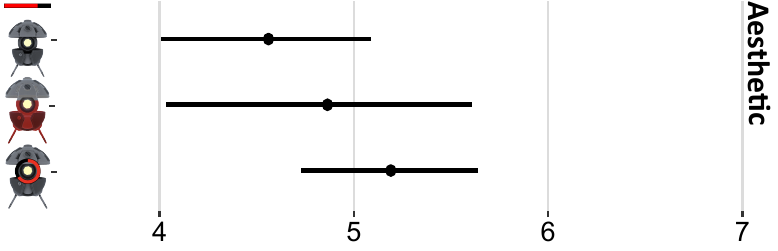}}&
    4.57&
    [4.02, 5.09]&
    \multirow{3}{*}{\includegraphics[height=\pictureheight]{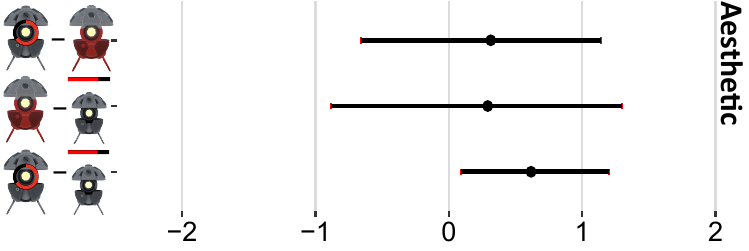}}&
    0.32&
    [-0.64, 1.14]
    \\
    &
    4.87&
    [4.04, 5.60]&
    &
    0.30&
    [-0.87, 1.30]
    \\
    &
    5.19&
    [4.73, 5.63]&
     &
    0.62&
    [0.10, 1.20]
    \\
    \specialrule{0em}{4ex}{0.1ex}
    \cline{1-6}
    \specialrule{0em}{2ex}{0.1ex}
    \multirow{3}{*}{\includegraphics[height=\pictureheight]{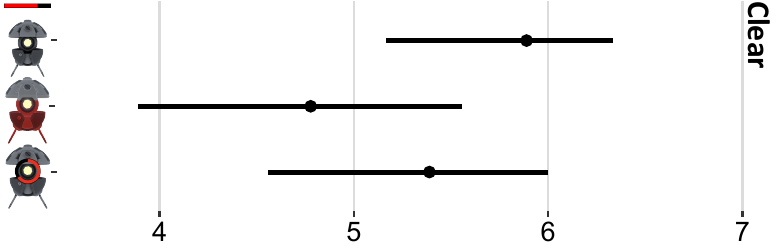}}&
    5.89&
    [5.17, 6.33]&
    \multirow{3}{*}{\includegraphics[height=\pictureheight]{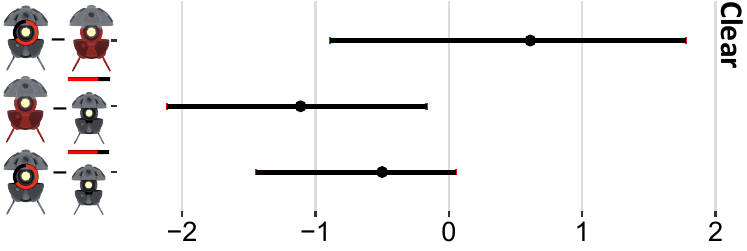}}&
    0.61&
    [-0.89, 1.78]
    \\
    &
    4.78&
    [3.89, 5.56]&
     &
    -1.11&
    [-2.11, -0.17]
    \\
    &
    5.39&
    [4.56, 6.00]&
     &
    -0.50&
    [-1.44, 0.06]
    \\
    \specialrule{0em}{0.1ex}{0.1ex}
    \\
    \multirow{3}{*}{\includegraphics[height=\pictureheight]{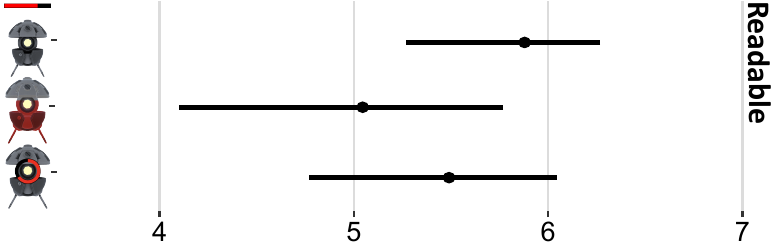}}&
    5.89&
    [5.28, 6.28]&
    \multirow{3}{*}{\includegraphics[height=\pictureheight]{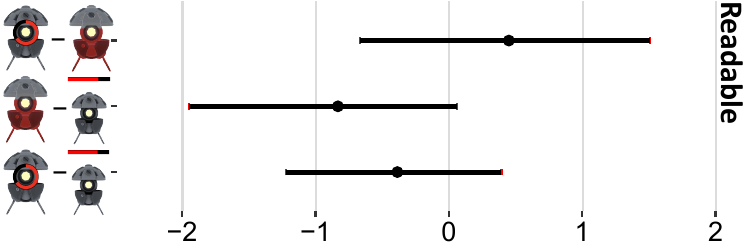}}&
    -0.44&
    [-0.67, 1.50]
    \\
    &
    5.05&
    [4.11, 5.78]&
    &
    -0.83&
    [-1.94, 0.06]
    \\
    &
    5.50&
    [4.78, 6.06]&
     &
    -0.39&
    [-1.22, 0.39]
    \\
    \specialrule{0em}{0.1ex}{0.1ex}
    \\
    \multirow{3}{*}{\includegraphics[height=\pictureheight]{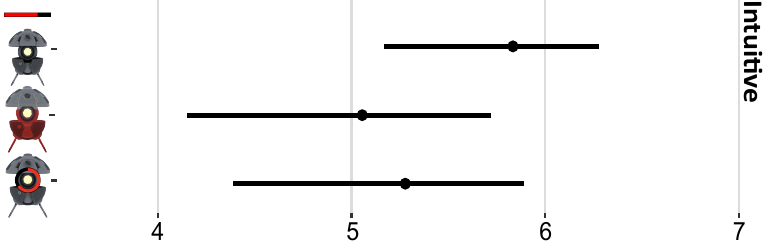}}&
    5.83&
    [5.17, 6.28]&
    \multirow{3}{*}{\includegraphics[height=\pictureheight]{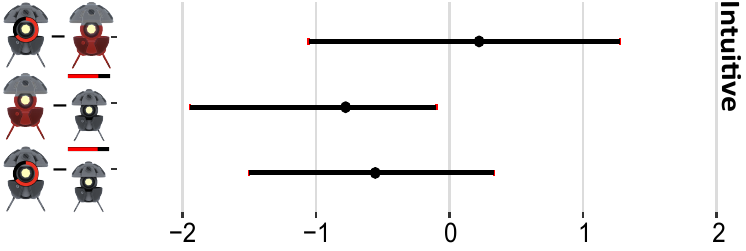}}&
    0.22&
    [-1.06, 1.28]
    \\
    &
    5.06&
    [4.15, 5.72]&
     &
    -0.78&
    [-1.94, -0.10]
    \\
    &
    5.28&
    [4.39, 5.89]&
    &
    -0.56&
    [-1.50, 0.33]
    \\
    \specialrule{0em}{0.1ex}{0.1ex}
    \\
    \multirow{3}{*}{\includegraphics[height=\pictureheight]{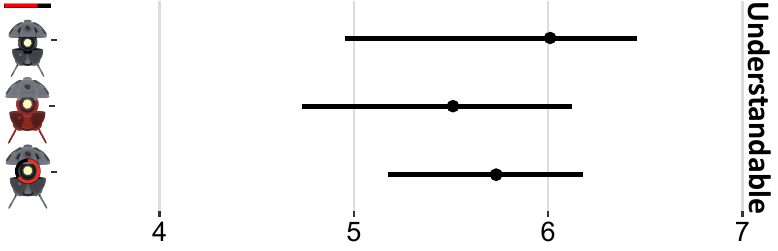}}&
    6.00&
    [4.94, 6.44]&
    \multirow{3}{*}{\includegraphics[height=\pictureheight]{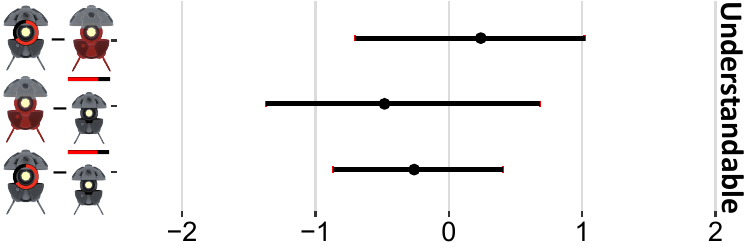}}&
    0.22&
    [-0.72, 1.00]
    \\
    &
    5.50&
    [4.72, 6.11]&
    &
    -0.50&
    [-1.39, 0.67]
    \\
    &
    5.72&
    [5.17, 6.17]&
    &
    -0.28&
    [-0.89, 0.39]
    \\
    \specialrule{0em}{4ex}{0.1ex}
    \centering Means and CIs & & & \centering Pairwise differences and CIs & & \\
    \end{tabular}
\end{table*}
\clearpage

\setlength{\picturewidth}{.4\linewidth}
\setlength{\pictureheight}{2.2cm}
\begin{table*}
\renewcommand{\arraystretch}{1.7}  
    \caption{Detailed results of participants' game performance. Top: Mean and pairwise differences between conditions of participants' killed robots. Middle: Mean and pairwise differences between conditions of participants' number of won and lost games. Bottom: Means and pairwise differences between conditions of participants' win and loss ratio (won or lost games over all games played).
    Error bars represent 95\% Bootstrap confidence intervals (CIs) in black, adjusted for pairwise comparisons with Bonferroni correction (in red). In the Pairwise difference image, CIs that do not cross 0 indicate evidence of a difference between the two techniques -  the tighter the CI of the difference and the furthest from 0, the stronger the evidence of a difference.  Our results do not provide clear evidence of a difference between designs, but there may be a trend for that the  \includegraphics[height = 0.32cm, valign=c]{Figures/icons/bar.pdf} \texttt{non-integrated design} to have a higher win ratio and lower loss ratio than the other two.}
    \label{tab:performanceStatisticalAnalysis}
    \scriptsize
    \begin{tabular}{@{}p{1.97\picturewidth}@{}
                    @{}r@{~}
                    @{}p{1.8cm}@{}
                    @{}p{1.87\picturewidth}@{}
                    @{}r@{~}
                    @{}p{1.5cm}@{}
                    }

    \multirow{3}{*}{\includegraphics[height=\pictureheight]{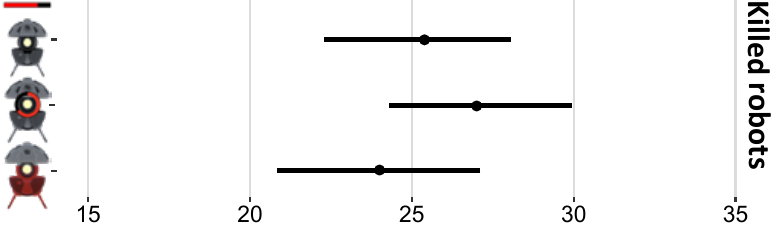}} & 
    25.39 & 
    [22.28, 28.06] & 
    \multirow{3}{*}{\includegraphics[height=\pictureheight]{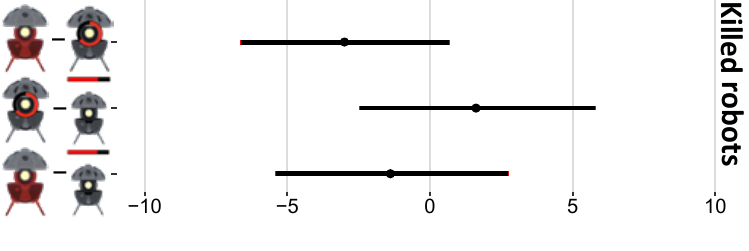}} & 
    -3.00 & 
    [-6.61, 0.67] 
    \\
    & 
    27.00 & 
    [24.28, 29.94] & 
    & 
    1.61 & 
    [-2.44, 5.78] 
    \\
    & 
    24.00 & 
    [20.83, 27.11] & 
    & 
    -1.39 & 
    [-5.39, 2.72] 
    \\
    
    \specialrule{0em}{4ex}{0.1ex}
    \cline{1-6}
    \specialrule{0em}{2ex}{0.1ex}

    \multirow{3}{*}{\includegraphics[height=\pictureheight]{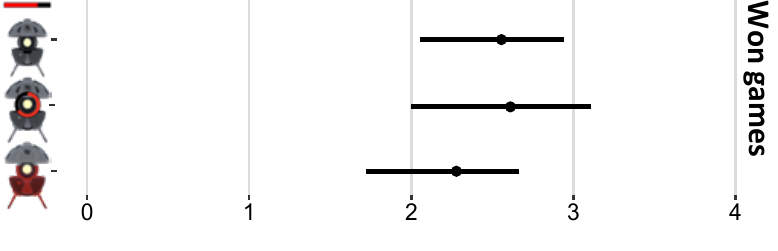}} & 
    2.56 & 
    [2.06, 2.94] & 
    \multirow{3}{*}{\includegraphics[height=\pictureheight]{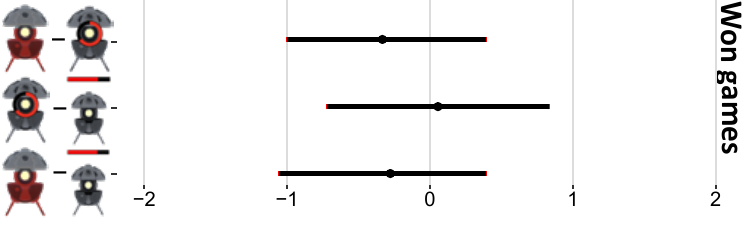}} & 
    -0.33 & 
    [-1.00, 0.39] 
    \\
    & 
    2.61 & 
    [2.00, 3.11] & 
    & 
    0.06 & 
    [-0.72, 0.83] 
    \\
    & 
    2.28 & 
    [1.72, 2.67] & 
    & 
    -0.28 & 
    [-1.06, 0.39] 
    \\

    \specialrule{0em}{4ex}{0.1ex}
    \specialrule{0em}{2ex}{0.1ex}
    
    \multirow{3}{*}{\includegraphics[height=\pictureheight]{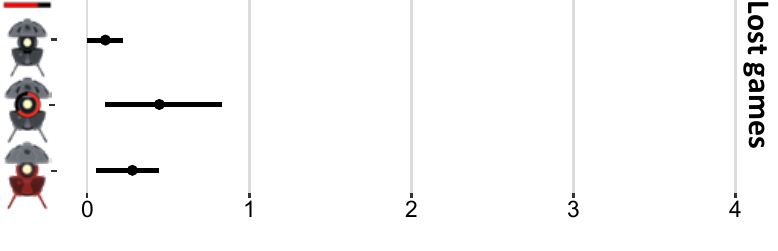}} & 
    0.11 & 
    [0.00, 0.22] & 
    \multirow{3}{*}{\includegraphics[height=\pictureheight]{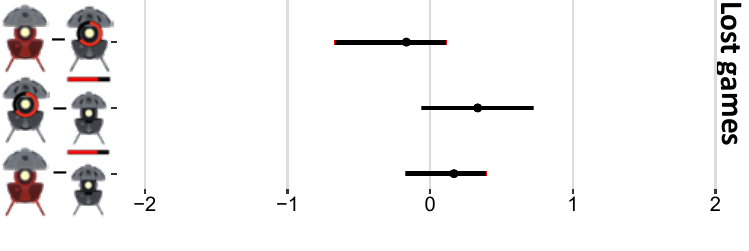}} & 
    -0.17 & 
    [-0.67, 0.11] 
    \\
    & 
    0.44 & 
    [0.11, 0.83] & 
    & 
    0.33 & 
    [-0.06, 0.72] 
    \\
    & 
    0.28 & 
    [0.06, 0.44] & 
    & 
    0.17 & 
    [-0.17, 0.39] 
    \\

    \specialrule{0em}{4ex}{0.1ex}
    \cline{1-6}
    \specialrule{0em}{2ex}{0.1ex}

    \multirow{3}{*}{\includegraphics[height=\pictureheight]{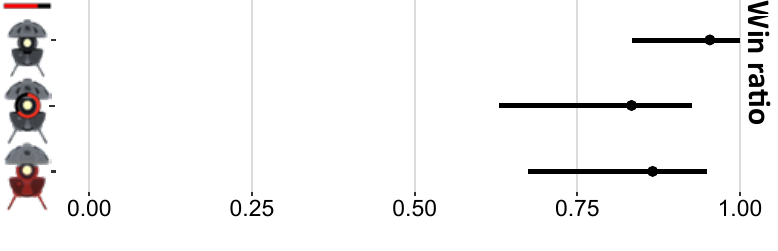}} & 
    0.95 & 
    [0.83, 1.00] & 
    \multirow{3}{*}{\includegraphics[height=\pictureheight]{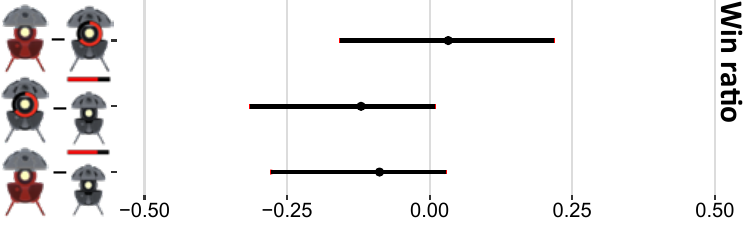}} & 
    0.03 & 
    [-0.16, 0.21] 
    \\
    & 
    0.83 & 
    [0.63, 0.93] & 
    & 
    -0.12 & 
    [-0.31, 0.01] 
    \\
    & 
    0.87 & 
    [0.67, 0.95] & 
    & 
    -0.09 & 
    [-0.28, 0.03] 
    \\
    
    \specialrule{0em}{4ex}{0.1ex}
    \specialrule{0em}{2ex}{0.1ex}

    \multirow{3}{*}{\includegraphics[height=\pictureheight]{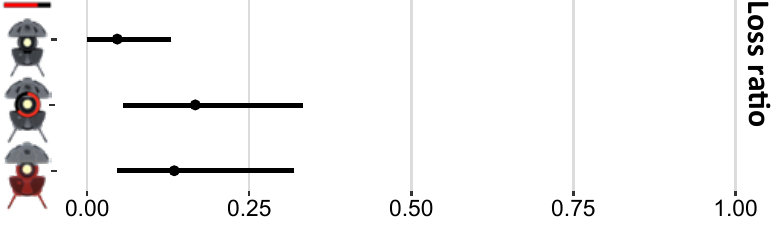}} & 
    0.05 & 
    [0.00, 0.13] & 
    \multirow{3}{*}{\includegraphics[height=\pictureheight]{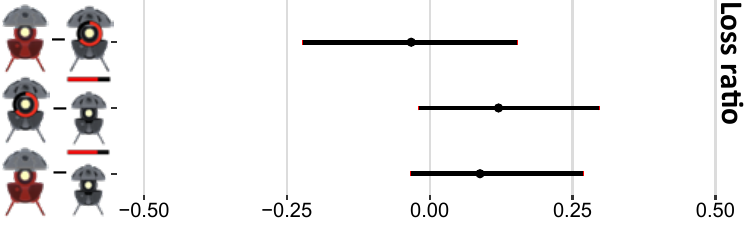}} & 
    -0.03 & 
    [-0.22, 0.15] 
    \\
    & 
    0.17 & 
    [0.06, 0.33] & 
    & 
    0.12 & 
    [-0.02, 0.30] 
    \\
    & 
    0.13 & 
    [0.05, 0.32] & 
    & 
    0.09 & 
    [-0.03, 0.27] 
    \\
   
    \specialrule{0em}{4ex}{0.1ex}
    \centering Means and CIs & & & \centering Pairwise differences and CIs & & \\
    \end{tabular}
\end{table*}

\end{document}